\documentclass[english,aps,pra,showpacs,floatfix,nofootinbib,twocolumn]{revtex4}
\usepackage{mathptmx}
\usepackage[latin9]{inputenc}
\usepackage{color}
\usepackage{babel}
\usepackage{amsmath}
\usepackage{amssymb}
\usepackage{graphicx}
\usepackage{subscript}
\usepackage[unicode=true,pdfusetitle,
 bookmarks=true,bookmarksnumbered=false,bookmarksopen=false,
 breaklinks=true,pdfborder={0 0 0},pdfborderstyle={},backref=false,colorlinks=true]
 {hyperref}
\usepackage{breakurl}

\makeatletter
\@ifundefined{textcolor}{}
{%
 \definecolor{BLACK}{gray}{0}
 \definecolor{WHITE}{gray}{1}
 \definecolor{RED}{rgb}{1,0,0}
 \definecolor{GREEN}{rgb}{0,1,0}
 \definecolor{BLUE}{rgb}{0,0,1}
 \definecolor{CYAN}{cmyk}{1,0,0,0}
 \definecolor{MAGENTA}{cmyk}{0,1,0,0}
 \definecolor{YELLOW}{cmyk}{0,0,1,0}
}

\@ifundefined{showcaptionsetup}{}{%
 \PassOptionsToPackage{caption=false}{subfig}}
\usepackage{subfig}
\makeatother

\begin{document}

\preprint{Version 0.3 \today}

\title{Near-field relaxation of a quantum emitter to 2D semiconductors:    
surface dissipation and exciton polaritons}

\author{Vasilios D. Karanikolas}
\email{karanikv@tcd.ie}

\affiliation{Photonics Group, School of Physics and CRANN,\\
 Trinity College Dublin, College Green, Dublin 2, Ireland}

\author{Cristian A. Marocico}

\affiliation{Photonics Group, School of Physics and CRANN,\\
 Trinity College Dublin, College Green, Dublin 2, Ireland}

\author{Paul R. Eastham}

\affiliation{Photonics Group, School of Physics and CRANN,\\
 Trinity College Dublin, College Green, Dublin 2, Ireland}

\author{A. Louise Bradley}

\affiliation{Photonics Group, School of Physics and CRANN,\\
 Trinity College Dublin, College Green, Dublin 2, Ireland}

\date{\today}

\begin{abstract}
The total spontaneous emission rate of a quantum emitter in the presence
of an infinite MoS\textsubscript{2} monolayer is enhanced by several
orders of magnitude, compared to its free-space value, due to the
excitation of surface exciton polariton modes and lossy modes. The
spectral and distance dependence of the spontaneous emission rate
are analyzed and the lossy-surface-wave, surface exciton polariton
mode and radiative contributions are identified. The transverse magnetic
and transverse electric exciton polariton modes can be excited for
different emission frequencies of the quantum emitter, and their contributions
to the total spontaneous emission rate are different. To calculate
these different decay rates, we use the non-Hermitian description
of light-matter interactions, employing a Green's tensor formalism.
The distance dependence follows different trends depending on the
emission energy of quantum emitter. For the case of the lossy surface
waves, the distance dependence follows a $z^{-n}$, $n=2,3,4$, trend.
When transverse magnetic exciton polariton modes are excited, they
dominate and characterize the distance dependence of the spontaneous
emission rate of a quantum emitter in the presence of the MoS\textsubscript{2}
layers. The interaction between a quantum emitter and a MoS\textsubscript{2}
superlattice is investigated and we observe a splitting of the modes
supported by the superlattice. Moreover, a blue shift of the peak
values of the spontaneous emission rate of a quantum emitter is observed
as the number of layers is increased. The field distribution profiles,
created by a quantum emitter, are used to explain this behavior.
\end{abstract}
\maketitle

\section{introduction\label{sec:Sec.I}}

The emission properties of quantum emitters are modified by their
environment \cite{Purcell1946}. In particular, through excitation
of surface plasmon modes, the spontaneous emission rate of a quantum
emitter (QE) can be enhanced by several orders of magnitude compared
with its free-space value \cite{Maier2005,Marocico2011}. Surface
plasmon polaritons are collective oscillations of electrons and the
electromagnetic field that are excited at the interface between a
dielectric and a conductor, and they are confined at this interface
and propagate along it. Noble metals, such as Au and Ag, are typically
used as plasmonic materials. However, the main disadvantage of using
noble metals is the fact that they have high losses in the optical
region of the spectrum \cite{Khurgin2015c}. As an alternative for
materials supporting surface plasmon modes, but with lower losses,
graphene can be considered \cite{Low2014,GarciadeAbajo2014}. Graphene
is a zero direct band-gap two-dimensional material of great potential
and with high mechanical capabilities \cite{Avouris2010}. However,
it also has a disadvantage, since it exhibits no plasmonic response
in the visible part of the spectrum, and acts only as a quencher when
interacting with quantum emitters emitting in the visible part of
the spectrum \cite{Gaudreau2013,Lee2014a}. 

In addition to surface plasmon modes, there are other surface modes
such as phonon and exciton polariton \cite{Yang1990,Caldwell2015}
modes. In previous years, a new family of two-dimensional materials,
the transition metal dichalconides (TMD), such as MoS\textsubscript{2},
SnS\textsubscript{2} and WeS\textsubscript{2}, have been subject
of intense theoretical \cite{Scholz2013,garstein2015,Khurgin2015}
and experimental investigations \cite{Britnell2013,Xia2014}. These
materials are direct bandgap semiconductors, with the conduction and
valence band edges at the doubly degenerate corners $(\pm\mathbf{K}\;\text{points})$
of the hexagonal Brilouin zone, and can have relatively high absorption
and intense photoluminescence \cite{Xiao2012,Yu2014}. We calculate
the spontaneous emission rate for a QE above a MoS$_{2}$ layer, and
find that the spontaneous emission rate is enhanced by many orders
of magnitude. We trace this effect to the near-field energy transfer
from the QE to the surface exciton polariton. In addition to implications
for energy transfer applications, such as photodetectors \cite{Lopez-Sanchez2013},
photovoltaic \cite{Kozawa2014} and light emitting devices \cite{Chakraborty2015,Amani2015,Messer2015},
our results show that low-dimensional materials can be used to study
polaritons and exciton-photon coupling phenomena without requiring
a microcavity \cite{Vasilevskiy2015}. Herein we demonstrate that
MoS\textsubscript{2} monolayers can support surface exciton polariton
modes and their influence on the optical properties of QEs is substantial. 

The interaction between quantum emitters (QEs) and multilayers of
TMD materials is of particular experimental interest. Many applications
can benefit from manipulating these interactions, such as photodetectors
\cite{Lopez-Sanchez2013}, electronic \cite{Radisavljevic2011}, photovoltaic
\cite{Kozawa2014} and light emitting devices \cite{Chakraborty2015,Amani2015,Messer2015}.
Investigating the spectral and distance dependence of the interactions
between QEs and TMD layers or monolayers is of absolute importance
for such applications. Various experimental studies have been performed
regarding the investigation of such interactions, and they report
contradicting results concerning the power law followed by the interaction
distance between the QE\textendash TMD layers, where different QEs
are considered for each case \cite{Prins2014,Kufer2015,Sampat2016,Zang2016,Chen,Goodfellow2016}.
A systematic analysis is needed to account for the spectral and distance
dependence of the QE\textendash TMD layer interaction. Here we focus
on material parameters describing the semiconducting behavior of MoS\textsubscript{2},
through the exciton energies and damping parameters \cite{Xiao2012,Zhang2014d,Vasilevskiy2015}. 

We find that transverse electric (TE) and transverse magnetic (TM)
exciton polariton modes are supported by a MoS\textsubscript{2} layer,
Fig.~\ref{fig:03}. The propagation length and penetration depth
of these modes are investigated. The SE rate of the QE is enhanced
several orders of magnitude for emission energies close to the exciton
energies, especially when the TM exciton polariton modes are excited,
in the presence of a single MoS\textsubscript{2} layer, see Fig.~\ref{fig:06}.
The different contributions to the total SE rate are presented: the
lossy surface wave (LSW), TE and TM exciton polariton modes, and radiative
emission contribution, for different QE\textendash MoS\textsubscript{2}
separations and emission energies of the QE. 

Additional physics appears as one goes beyond the single layer structure
to multilayers. In particular, we show that the electromagnetic coupling
between the layers splits the degeneracy of the exciton polariton
modes, even in the absence of direct electronic coupling. We find,
Fig.~\ref{fig:07}, that the electromagnetic coupling between the
layers leads to a blue-shift in the peak of the spontaneous emission
rate with increasing number of layers. This may provide an explanation
for the apparent different dependences of the emission rate with the
layer number observed in experiments \cite{Chen,Zang2016}.

In Sec.~\ref{sec:Sec.II} we introduce the mathematical method for
studying the QE\textendash MoS\textsubscript{2} structure. The QE
is described as a two-level system and the Green's tensor formalism
is used to describe the light-matter interaction in the non-Hermitian
description of quantum electrodynamics, \ref{sec:Sec.IIa}. The optical
response of the MoS\textsubscript{2} layer is modeled by the surface
conductivity, Sec.~\ref{sec:Sec.IIb}. In Sec.~\ref{sec:Sec.III}
we give the results. We start in Sec.~\ref{sec:Sec.IIIA} by analyzing
the surface exciton polariton when a single exciton resonance is considered
in the surface conductivity. When two exciton resonances are considered,
we see that two bands are formed, corresponding to the TE and TM exciton
polariton modes, Sec.~\ref{sec:Sec.IIIA}. The propagation length
and penetration depth of the TE and TM exciton polariton modes are
analyzed. In Sec.~\ref{sec:Sec.IIIB}, the interaction between a
QE and a free-standing MoS\textsubscript{2} layer is considered.
The spectral and distance dependence is analyzed and the different
contributions are studied. The LSW, TE and TM exciton polariton modes
and radiative emission contributions to the SE rate of a QE, at different
positions and emission energies, are presented. In Sec.~\ref{sec:Sec.IIIC}
we focus on the interaction between a QE and MoS\textsubscript{2}
planar superlattices. We observe that the TE and TM exciton polariton
modes bands are still split and that multibands are also formed, due
to interlayer scattering. The SE peak of the QE is blue-shifted and
the absolute value of the SE rate enhancement decreases. Finally,
in Sec.~\ref{sec:IV} we give some concluding results and future
steps for research in the field.

\section{Mathematical methods\label{sec:Sec.II}}

\subsection{Spontaneous emission rate\label{sec:Sec.IIa}}

The quantum emitters (QEs) considered in this paper are approximated
as two-level systems. Various emitters, such as atoms, molecules,
quantum dots and NV color centers, can be approximated in this way.
The ground state of the QE is denoted as $|g\rangle$, and the excited
state as $|e\rangle$. The transition frequencies from the excited
to the ground state and the transition dipole matrix element are denoted
as $\omega_{\text{T}}$ and $\boldsymbol{\mu}$, respectively. The
multipolar Hamiltonian is used to describe a QE interacting with the
electromagnetic field \cite{Dung1998,Dung2000}, and it has the form
\begin{align}
\hat{H}=\hat{H}_{\textrm{em}}= & \int\textrm{d}^{3}r\int\limits _{0}^{\infty}\textrm{d}\omega\,\hbar\omega\,\hat{\mathbf{f}}^{\dagger}(\mathbf{r},\omega)\cdot\hat{\mathbf{f}}(\mathbf{r},\omega)+\hbar\omega_{T}\sigma^{\dagger}\sigma\nonumber \\
 & -\int d\omega\big[\hat{\boldsymbol{\mu}}\cdot\mathbf{\hat{E}}(\mathbf{r},\omega)+H.c.\big],\label{eq:01}
\end{align}
where $\hat{\boldsymbol{\mu}}=\boldsymbol{\mu}\sigma^{\dagger}+\boldsymbol{\mu}^{*}\sigma^{-}$
is the transition dipole operator of the two level system, with $\boldsymbol{\mu}$
being the transition dipole moment of the system between its ground
and excited states. The electric field operator has the form
\begin{equation}
\hat{\mathbf{E}}(\mathbf{r},\omega)=i\sqrt{\frac{\hbar}{\pi\varepsilon_{0}}}\frac{\omega^{2}}{c^{2}}\int\textrm{d}^{3}s\,\sqrt{\varepsilon^{\prime\prime}(\mathbf{s},\omega)}\,\mathfrak{G}(\mathbf{r},\mathbf{s},\omega)\cdot\hat{\mathbf{f}}(\mathbf{s},\omega),\label{eq:02}
\end{equation}
where $\hat{\mathbf{f}}(\mathbf{s},\omega)$ and $\hat{\mathbf{f}}^{\dagger}(\mathbf{s},\omega)$
are creation and annihilation operators for medium-dressed states,
which account for the various modes provided by the environment, such
as the LSWs, surface exciton polariton and radiative modes considered
in this paper.

An excited quantum emitter interacts with its environment through
the electromagnetic field and relaxes from its excited state to the
ground state by emitting a photon or exciting any of the dressed states
supported by its environment. The initial state of the system is denoted
as $|i\rangle=|e\rangle\otimes|0\rangle$, where the QE is in the
excited state and the electromagnetic field is in its vacuum state.
The quantum emitter will not stay indefinitely excited, but will relax
to the medium dressed states and therefore the EM field will be in
a $|1(\mathbf{k},p)\rangle=\hat{f}_{i}^{\dagger}(\mathbf{r},\omega)|0\rangle$
state; $p$ and $\mathbf{k}$ are the polarization and wavevector,
respectively. The final state of the entire system therefore has the
form $|f\rangle=|g\rangle\otimes\hat{f}_{i}^{\dagger}(\mathbf{r},\omega)|0\rangle$.
By applying Fermi's golden rule and summing over all final states,
the expression for the SE rate $\Gamma$ is obtained as:
\begin{equation}
\Gamma(\mathbf{r},\omega)=\frac{2\omega^{2}\mu^{2}}{\hbar\varepsilon_{0}c^{2}}\hat{\mathbf{n}}\cdot\textrm{Im}\,\mathfrak{G}(\mathbf{r},\mathbf{r},\omega)\cdot\hat{\mathbf{n}},\label{eq:03}
\end{equation}
where $\mathbf{\hat{n}}$ is a unit vector along the direction of
the transition dipole moment, $\boldsymbol{\mu}$, and $\mathfrak{G}(\mathbf{r},\mathbf{s},\omega)$
is the Green's tensor representing the response of the geometry under
consideration to a point-like excitation. In order to quantify the
influence of the environment on the QE emission, the normalized SE
rate is defined as:

\begin{equation}
\tilde{\Gamma}=\frac{\Gamma}{\Gamma_{0}}=\sqrt{\varepsilon}+\frac{6\pi c}{\omega}\hat{n}_{i}\mathrm{\text{Im}}\,\mathfrak{G}_{\text{S}}^{ii}(\mathbf{r},\mathbf{r},\omega)\hat{n}_{i},\label{eq:04}
\end{equation}
where $\varepsilon$ is the permittivity of the host medium, $\Gamma_{0}$
is given by the Einstein $A$-coefficient $\Gamma_{0}=\omega^{3}\mu^{2}/3\pi c^{3}\hbar\varepsilon_{0}$
and $\mathfrak{G}_{\text{S}}$ is the scattering part of the Green's
tensor calculated at the QE position $\mathbf{r}$.

The normalized SE rate for the $x$ and $z$ orientations of the transition
dipole moment of a QE in the presence of an infinite MoS\textsubscript{2}
layer are given by the expressions\begin{subequations}\label{eq:05}
\begin{equation}
\tilde{\Gamma}_{z}=\sqrt{\varepsilon_{1}}+\frac{3c}{2\omega}\text{Im}\Bigg(i\int\limits _{0}^{\infty}\text{d}k_{s}\frac{k_{s}^{3}}{k_{z1}k_{1}^{2}}R_{N}^{11}e^{2ik_{z1}z}\Bigg),\label{eq:05a}
\end{equation}
\begin{equation}
\tilde{\Gamma}_{x}=\sqrt{\varepsilon_{1}}+\frac{3c}{4\omega}\text{Im}\left[i\int\limits _{0}^{\infty}\text{d}k_{s}\frac{k_{s}}{k_{1}}\left(R_{M}^{11}+\frac{k_{z1}^{2}}{k_{1}^{2}}R_{N}^{11}\right)e^{2ik_{z1}z}\right].\label{eq:05b}
\end{equation}
\end{subequations}More details on the calculation of the Green's
tensor, when an infinite MoS\textsubscript{2} layer and superlattice
are considered as the environment of a QE, are given in App.~\ref{sec:AppA}. 

Here $R_{N}$ and $R_{M}$ are Fresnel coefficients for the reflection
from the surface, defined in Appendix \ref{sec:AppA}. For a single
free-standing layer $(\varepsilon_{1}=\varepsilon_{2}=1)$, with surface
conductivity $\sigma$ they are \cite{Hanson2013,Nikitin2013}\begin{subequations}\label{eqs:06}
\begin{equation}
R_{M}^{11}=\frac{-\alpha k_{0}}{k_{z}+\alpha k_{0}},\quad R_{N}^{11}=\frac{\alpha k_{z}}{k_{0}+\alpha k_{z}}\label{eq:06a}
\end{equation}
\begin{equation}
R_{M}^{21}=\frac{k_{z}}{k_{z}+\alpha k_{0}},\quad R_{N}^{21}=\frac{k_{0}}{k_{0}+\alpha k_{z}},\label{eq:06b}
\end{equation}
where $\alpha=2\pi\sigma/c$ and $k_{z}=\sqrt{k_{0}^{2}-k_{s}^{2}}$.\end{subequations} 

\subsection{Surface conductivity\label{sec:Sec.IIb}}

\begin{figure}[t]
\includegraphics[width=0.9\columnwidth]{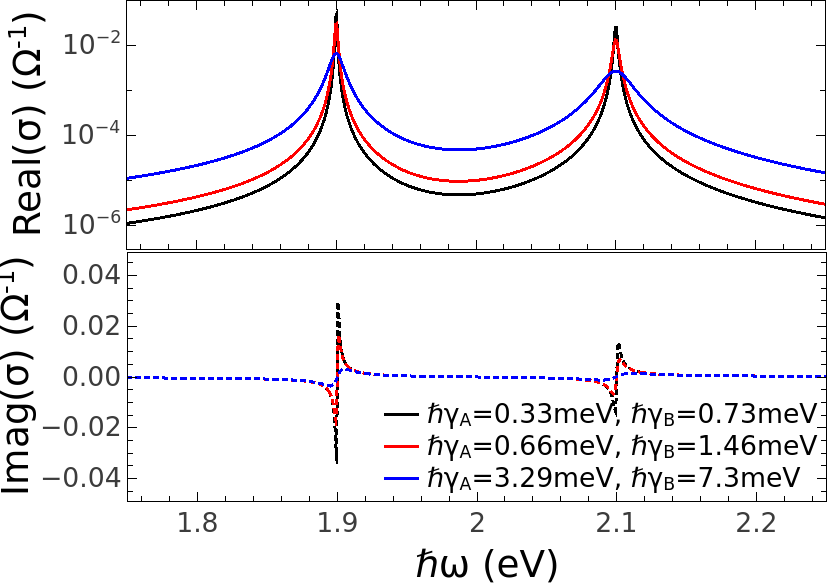}\caption{(Color online) Real and imaginary parts of the surface conductivity
of MoS\protect\textsubscript{2}, $\sigma_{\text{res}}$, given by
Eq.~\eqref{eq:07} for different values of the damping parameters,
$\hbar\gamma_{A}=0.3\,\text{meV},\,0.7\,\text{meV}$ and $3.3\,\text{meV}$
and $\hbar\gamma_{B}=0.7\,\text{meV},\,1.4\,\text{meV}$ and $7.3\,\text{meV}$.\label{fig:01}}
\end{figure}
MoS\textsubscript{2} is a direct gap semiconductor with relatively
intense photoluminescence \cite{Amani2015}. The resonance part of
the 2-dimensional optical conductivity of the MoS\textsubscript{2},
$\sigma_{\text{res}}$, takes into account the interaction of light
with the lowest energy $A$ and $B$ excitons and is given by
\begin{equation}
\sigma_{\text{res}}(\omega)=\frac{4\alpha_{0}\hbar cv^{2}}{\pi a_{\text{ex}}^{2}\omega}\sum_{k=A,B}\frac{-i}{E_{k}-\hbar\omega-i\hbar\gamma_{k}},\label{eq:07}
\end{equation}
where $\alpha_{0}$ is the fine structure constant, $a_{\text{ex}}=0.8\,\text{nm}$
is the exciton Bohr radius, the damping parameters are $\gamma_{A}$
and $\gamma_{B}$, and the exciton energies are $E_{A}=1.9\,\text{eV}$
and $E_{B}=2.1\,\text{eV}$. $v$ is a constant velocity, which is
connected with the hopping parameter, and for MoS\textsubscript{2}
we use the value $v=0.55\,\text{nm/fs}$ \cite{Xiao2012,Zhang2014d}.
In Fig.~\ref{fig:01} we present the real and imaginary parts of
the surface conductivity for different values of the damping parameters,
$\gamma_{A}$ and $\gamma_{B}$\cite{Vasilevskiy2015}. The damping
parameters, $\gamma_{\text{A}}$ and $\gamma_{\text{B}}$, are connected
with the quality of the MoS\textsubscript{2} layer at different temperatures,
and for that reason we choose to investigate a broader spectrum of
parameters to account for the different mechanism of losses \cite{Palummo2015}.
The real part of the surface conductivity, $\sigma_{\text{res}}\left(\omega\right)$,
is connected with the losses, the higher its values, the more lossy
the material. We observe in Fig.~\ref{fig:01} that as the value
of the damping parameters increases, the peaks of the real part of
the surface conductivity in Fig.~\ref{fig:01} become broader. At
the exciton energies, $E_{A}$ and $E_{B}$, the losses are largest
for the smallest value of the damping parameters, $\gamma_{A}$ and
$\gamma_{B}$, because they give the linewidth of the resonance, but
away from them the real part of $\sigma_{\text{res}}$ increases as
the damping increases. The sign of the imaginary part of the surface
conductivity, $\sigma_{\text{res}}(\omega)$, determines the type
of modes supported by the MoS\textsubscript{2} layer and how dispersive
they are. More details on this will be given in the next section.

At even higher energies, the interband transitions need to be included
in the model describing the surface conductivity. We model these transitions
with an expression of the form\begin{widetext}
\begin{equation}
\text{\text{Real}}\left(\sigma_{\text{inter}}\right)=\frac{m\sigma_{0}\theta(\omega-\omega_{B})}{\sqrt{1+2E_{B}\beta+\Omega^{2}}}\left[1+\frac{1+2E_{B}\beta}{\Omega^{2}}\left(1+E_{B}\beta-\sqrt{1+2E_{B}\beta+\Omega^{2}}\right)\right],\label{eq:08}
\end{equation}
\end{widetext}where $\hbar\omega_{B}=E_{B}$, $\Omega=\hbar\omega/E_{B}$
and $\beta$ is a mixing parameter, for MoS\textsubscript{2} $E_{B}\beta=0.84$
\cite{Stauber2015a}. The parameter $m$ is for scaling the absorption
described by Eq.~\eqref{eq:08}. As we will see in Sec.~\ref{sec:Sec.IIIA},
the excitonic effects described by Eq.~\eqref{eq:08} are not important
in the energy spectrum we focus our analysis on, in particular for
energies close to the exciton resonances $E_{A}$ and $E_{B}$, $1.7\,\text{eV}<\hbar\omega<2.2\,\text{eV}$.

The emphasis of this paper is to theoretically investigate light-matter
interactions, thus, we choose to use a theoretical expression to describe
the optical response of the MoS\textsubscript{2}. This is done in
order to keep the discussion as general as possible. We choose the
material parameters connected with MoS\textsubscript{2}, and these
can be easily modified to study the interaction between a QE and any
TMD superlattice or thin semiconducting quantum well. Furthermore,
the material parameters for the MoS\textsubscript{2} are connected
with the quality of a specific sample, and thus experimentally, they
vary from study to study.

\section{Results\label{sec:Sec.III}}

\subsection{Surface exciton polariton modes\label{sec:Sec.IIIA}}

\begin{figure}
\includegraphics[width=0.45\textwidth]{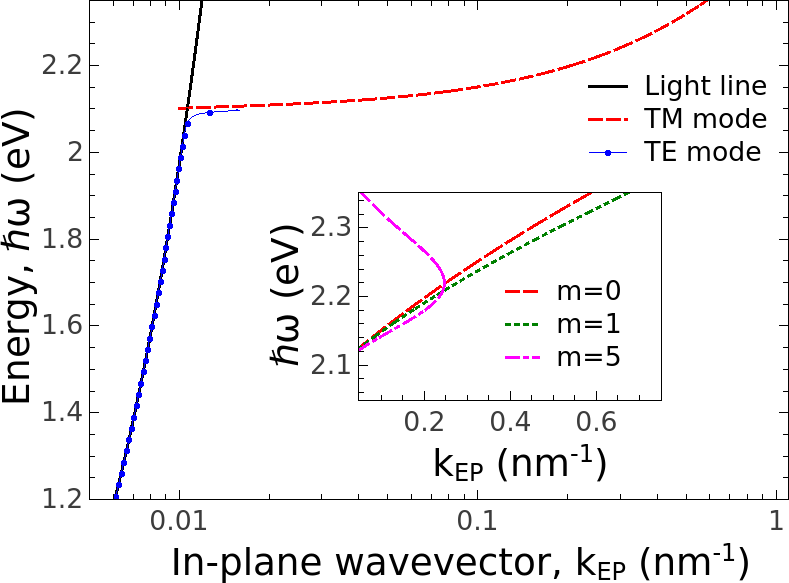}\caption{(Color online) Dispersion relation, $\text{Re}(k_{\text{EP}}(\omega))$,
for a 2D free-standing material, when a single exciton is considered,
$E_{B}=2.1\,\text{eV}$. The value of the damping parameter is $\gamma_{B}=0.7\,\text{meV}$.
In the inset the dispersion relation is presented when the interband
transitions are considered, Eq.~\ref{fig:08a}.\label{fig:02}}
\end{figure}
We start our analysis by considering a single exciton polariton mode
shown in Fig.~\ref{fig:02}, supported by a two-dimensional material.
The exciton energy is $E_{B}=2.1\,\text{eV}$ and the damping parameter
has a value of $\hbar\gamma_{B}=0.7\,\text{meV}$. We use exactly
the same parameters as when describing the MoS\textsubscript{2} layer,
only ignoring the exciton with energy $E_{A}$. In Fig.~\ref{fig:02}
the band structure of the transverse electric (TE) and transverse
magnetic (TM) exciton polariton modes for a free standing ($\varepsilon_{1}=\varepsilon_{2}=1$)
single exciton layer are shown. The dispersion relation of the TE
exciton polariton mode is calculated by setting the denominator of
$R_{M}^{11}$, Eq.~\eqref{eq:06a}, to zero which gives
\begin{equation}
k_{\text{EP}}^{\text{TE}}=\frac{\omega}{c}\sqrt{1-4\pi^{2}\sigma_{\text{res}}^{2}/c^{2}}.\label{eq:09}
\end{equation}
The TE mode is supported by the MoS\textsubscript{2} layer only when
the $\text{Im}(\sigma_{\text{res}})<0$, whereas for $\text{Im}(\sigma_{\text{res}})>0$
the TE mode is on the improper Riemann sheet \cite{Mikhailov2007a}.
$k_{\text{EP}}^{\text{TE}}$ is the in-plane wave vector of the TE
exciton polariton mode propagating on the MoS\textsubscript{2} layer.
The available TM modes are found similarly by setting the denominator
of the reflection coefficient $R_{N}^{11}$, Eq.~\eqref{eq:06a},
to zero and we get the expression 
\begin{equation}
k_{\text{EP}}^{\text{TM}}=\frac{\omega}{c}\sqrt{1-c^{2}/4\pi^{2}\sigma_{\text{res}}^{2}},\label{eq:10}
\end{equation}
which gives the dispersion relation $k_{\text{EP}}^{\text{TM}}(\omega)$,
the relation between the frequency, $\omega,$ and the TM exciton
polariton mode in-plane wave vector, $k_{\text{EP}}^{\text{TM}}$.
The TM exciton polariton modes can propagate on the MoS\textsubscript{2}
layer only when $\text{Im}(\sigma_{\text{res}})>0$, whereas when
$\text{Im}(\sigma_{\text{res}})<0$ the TM mode given by Eq.~\eqref{eq:10}
is on the improper Riemann sheet \cite{Hanson2015,Karanikolas2015}.

For the case of a single exciton, the imaginary part of the surface
conductivity is negative for energies below the exciton energy, $\hbar\omega<2.1\,\text{eV}$,
thus allowing only TE exciton polariton modes to propagate. On the
other hand for $\hbar\omega>2.1\,\text{eV}$, above the exciton energy
$E_{B}$, only TM exciton polariton modes are supported. As we observe
in Fig.~\ref{fig:02}, the TE modes are very close to the light-line,
which means that these modes are loosely confined to the MoS\textsubscript{2}
layer. It is only very close to the exciton energy, $E_{B}$, that
they start to become dispersive. The TM modes are clearly more dispersive
and they are tightly confined to the MoS\textsubscript{2} layer.

In the inset of Fig.~\ref{fig:02} the TM exciton polariton mode
is presented for energies $\hbar\omega>E_{B}$, in the case when interband
transitions are also included, Eq.~\eqref{eq:08}. We consider the
case of $m=0,\,1$ and $5$ in Eq.~\eqref{eq:08}. We observe that
as the value of $m$ is increased, the dispersion relation, $\text{Re}\left(k_{\text{EP}}^{\text{TM}}(\omega)\right)$,
starts to bend back, towards the light line, for high energies, $\hbar\omega\gtrsim2.2\,\text{eV}$.
This is due to the higher losses caused by electron-hole pair generation.
The dispersion relation has similar behavior to noble metal thin films
at higher energies \cite{Dionne2005}. We observe that at energies
up to $2.2\,\text{eV}$, the dispersion lines, for the different values
of $m$, are very close. For that reason, in the rest of this paper
we ignore the effect of the interband transitions, which are small
in the energy range we investigate, $1.7\,\text{eV}<\hbar\omega<2.2\,\text{eV}$.

\begin{figure*}
\subfloat[\label{fig:03a}]{\centering{}\includegraphics[width=0.32\textwidth]{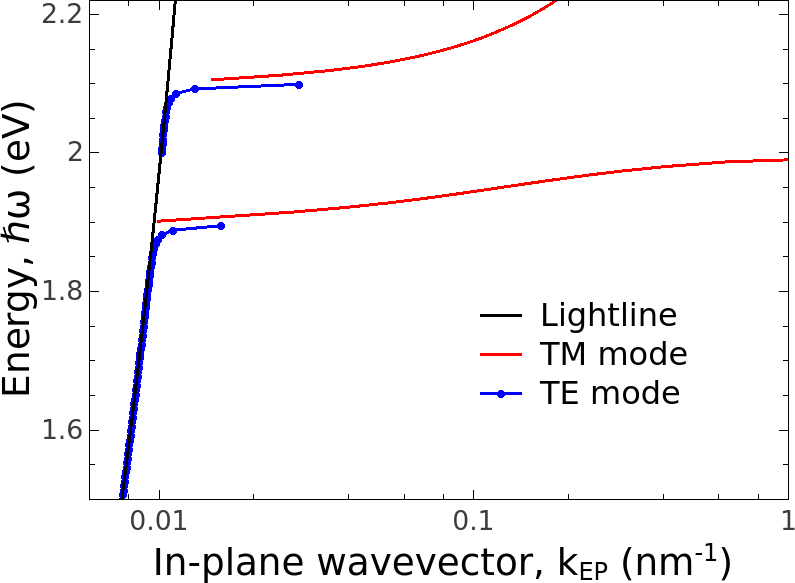}}~~~\subfloat[\label{fig:03b}]{\centering{}\includegraphics[width=0.32\textwidth]{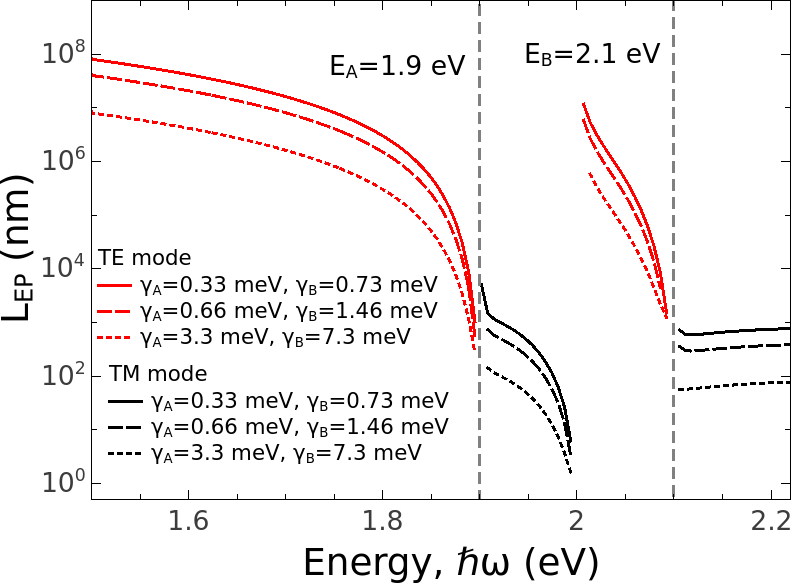}}~~~\subfloat[\label{fig:03c}]{\centering{}\includegraphics[width=0.32\textwidth]{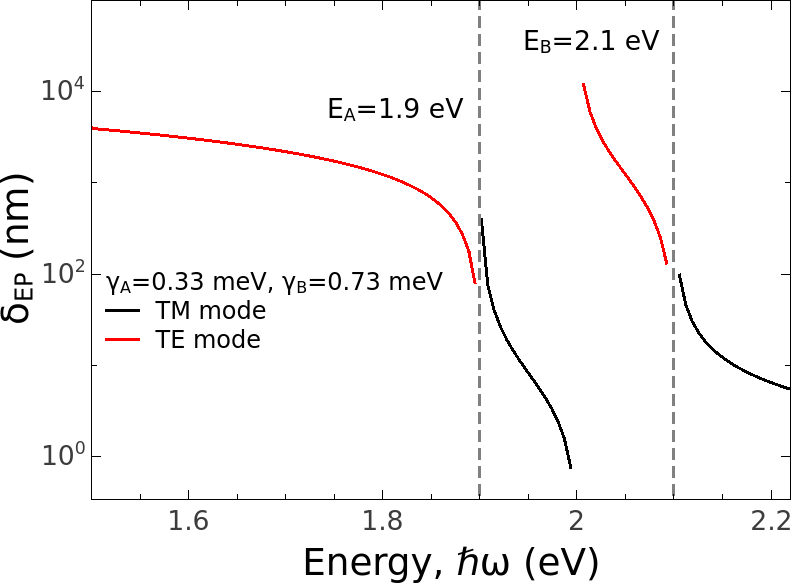}}\caption{(Color online) (a) Plot of the dispersion relation, $\text{Re}(k_{\text{EP}}(\omega))$,
when a free standing MoS$_{2}$ monolayer, $\varepsilon_{1}=\varepsilon_{2}=1$,
is considered. (b-c) Plot of the propagation length, $L_{\text{EP}}$,
and the penetration depth, $\delta_{\text{EP}}$, respectively. Different
values of the damping parameters, $\gamma_{A}$ and $\gamma_{B}$,
are considered. More details in the legends. \label{fig:03}}
\end{figure*}
We turn the discussion now to the case in which we consider both excitons
in the surface conductivity, $\sigma_{\text{res}}$, which matches
the physical material parameters of MoS\textsubscript{2}, Eq.~\eqref{eq:07}.
In Fig.~\ref{fig:03a} we present a plot of the dispersion relation,
$\text{Re}\left(k_{\text{EP}}(\omega)\right)$, for the TE and TM
exciton polariton modes. Due to the presence of two excitons with
energies $E_{A}=1.9\,\text{eV}$ and $E_{B}=2.1\,\text{eV}$, the
TE and TM exciton polariton modes split into two branches \cite{Agranovich1997}.
Again, an analogy can be drawn with the case of a metallic thin film,
which is sandwiched between two materials with different dielectric
permittivities \cite{Dionne2005}. Two surface plasmon polariton modes
are present in this case, due to the two different metal-dielectric
interfaces. In our case, the presence of the two excitons, with close
energies, is the reason for the dispersion relation in Fig.~\ref{fig:03}.
It implies that there is a change of sign for $\text{Im}\left(\sigma_{\text{res}}\right)$,
Fig.~\ref{fig:01}. In particular, for energies $\hbar\omega<E_{A}$
the imaginary part of the surface conductivity is negative $\text{Im}\left(\sigma_{\text{res}}\right)<0$,
thus only TE exciton polariton modes are supported, Fig.~\ref{fig:03a}.
These modes lie very close to the light line and are only loosely
confined to the MoS\textsubscript{2} layer and only very close to
the exciton energy $E_{A}$ do they become more dispersive. At energies
$E_{A}<\hbar\omega<2\,\text{eV}$ the $\text{Im}\left(\sigma_{\text{res}}\right)>0$,
thus TM exciton polariton modes are supported which are highly dispersive
and the value of the in-plane wave vector, $k_{\text{EP}}^{\text{TM}}$,
is larger by up to two orders of magnitude than the free-space wavevector,
$k_{0}=\omega/c$. At the energy of $\hbar\omega=2\,\text{eV}$, $\text{Im}\left(\sigma_{\text{res}}\right)\thickapprox0$
and at this point the imaginary part of the surface conductivity now
changes sign from plus to minus, due to the interaction between the
two exciton resonances; thus, for energies $2\,\text{eV}<\hbar\omega<E_{B}$,
we have $\text{Im}\left(\sigma_{\text{res}}\right)<0$, and TE exciton
polariton modes are supported. For $\hbar\omega>E_{B}$, $\text{Im}\left(\sigma_{\text{res}}\right)>0$
and TM exciton polariton modes are again supported. In Fig.~\ref{fig:03a}
we consider as damping parameters the values $\hbar\gamma_{A}=0.33\,\text{meV}$
and $\hbar\gamma_{B}=0.70\,\text{meV}$, and increasing these values
shows small influence on the real part of the in-plane wavevector
of the exciton polariton mode, $k_{\text{EP}}$. $\lambda_{\text{EP}}=2\pi/k_{\text{EP}}$
gives the propagation wavelength of the exciton polariton mode.

The imaginary part of the in-plane wavevector, $k_{\text{EP}}$, is
connected with the propagation length of the exciton polariton mode
$L_{\text{EP}}=1/\text{Im}\left(k_{\text{EP}}\right)$. In Fig.~\ref{fig:03b}
the propagation length, $L_{\text{EP}}$, is shown as a function of
energy for different values of the damping parameters, $\gamma_{A}$
and $\gamma_{B}$ (see the legend of Fig.~\ref{fig:03b} for more
details). We again observe the different intervals where the TE or
TM exciton polariton modes are excited, depending on the sign of $\text{Im}\left(\sigma_{\text{res}}\right)$.
As we have already pointed out, the TE modes are loosely confined
to the MoS\textsubscript{2} layer, thus their propagation length
is very large and its value differs from the TM exciton polariton
modes propagation length by up to 6 orders of magnitude. As the damping
is increased, the propagation length, $L_{\text{EP}}$, decreases.
It is also seen that the real part of the surface conductivity, $\sigma_{\text{res}},$
which is connected with the material losses, increases at energies
of the exciton resonances with increasing damping, Fig.~\ref{fig:01}.

The penetration depth is defined as $\delta_{\text{EP}}=1/\text{Im}\left(k_{z}^{\text{EP}}\right)$,
where $k_{z}^{\text{EP}}=\sqrt{k_{0}^{2}-k_{\text{EP}}^{2}}$, and
is connected with the extent of the exciton polariton mode in the
direction perpendicular to the MoS\textsubscript{2} layer. In Fig.~\ref{fig:03c}
the penetration depth is presented as a function of energy with damping
parameters $\hbar\gamma_{A}=0.33\,\text{meV}$ and $\hbar\gamma_{B}=0.70\,\text{meV}$.
Again, the sign of $\text{Im}\left(\sigma_{\text{res}}\right)$ gives
the different intervals where the TE and TM exciton polariton modes
propagate. The TE modes are only loosely confined to the MoS\textsubscript{2}
layer, are essentially radiative modes, as we will see in the next
section, and have only a small contribution to the modification of
the emission properties of a QE in proximity to the MoS$_{2}$ layer. 

\subsection{Spontaneous emission in the presence of a single MoS\protect\textsubscript{2}
layer\label{sec:Sec.IIIB}}

\begin{figure*}
\subfloat[\label{fig:04a}]{\includegraphics[width=0.32\textwidth]{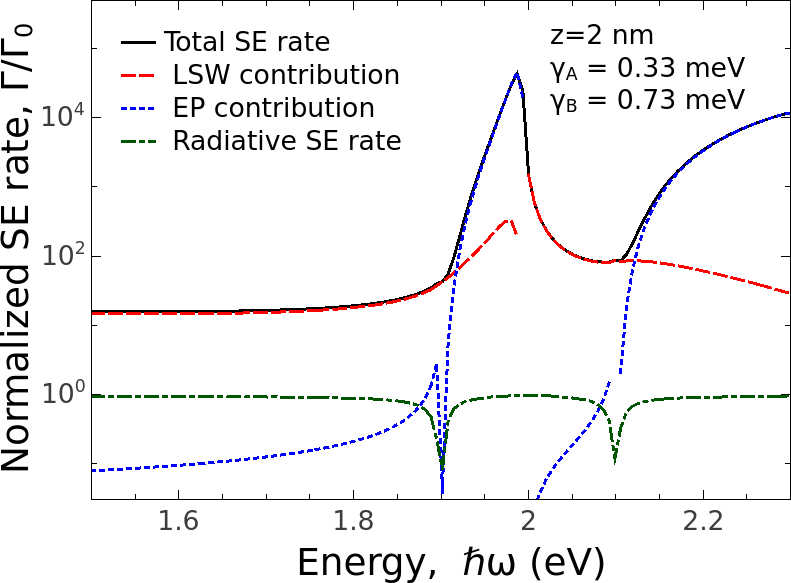}

}~~~\subfloat[\label{fig:04b}]{\includegraphics[width=0.32\textwidth]{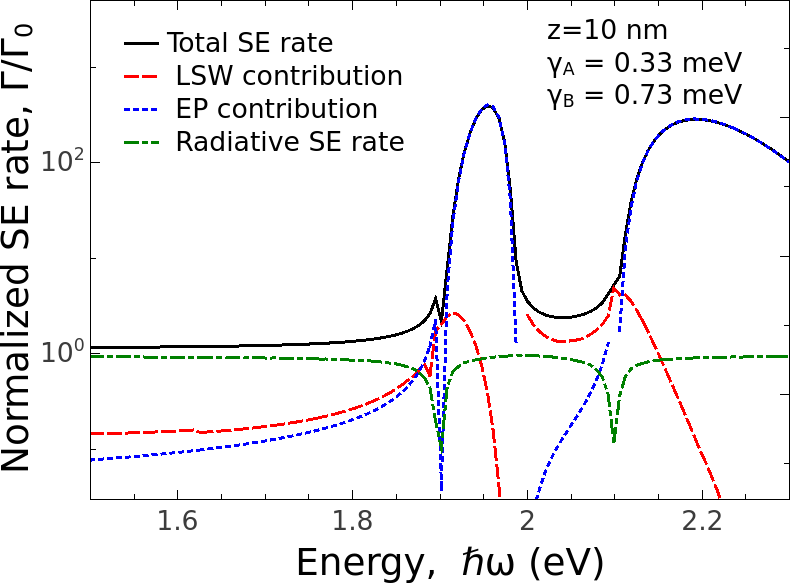}

}

\subfloat[\label{fig:04c}]{\includegraphics[width=0.32\textwidth]{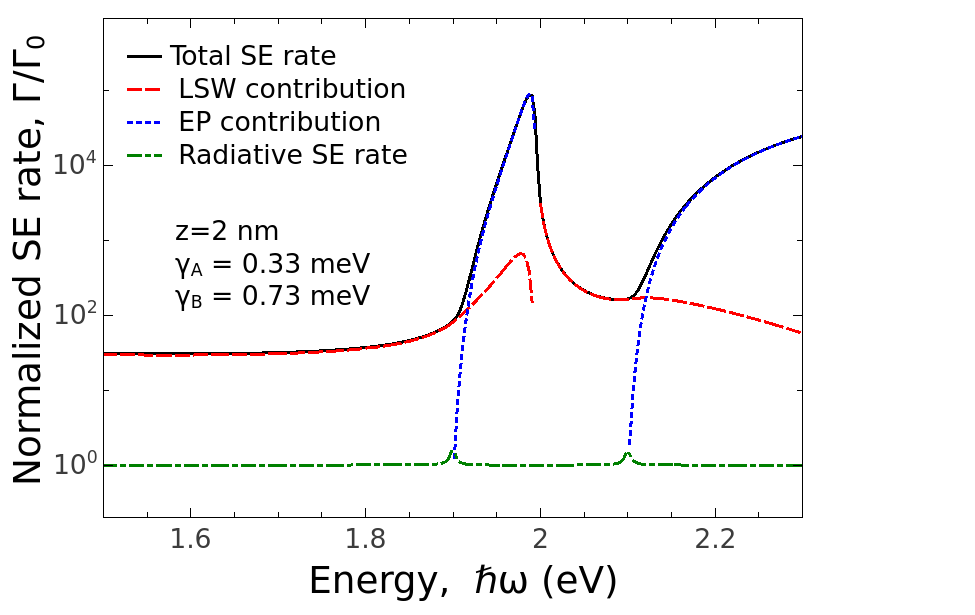}

}~~~\subfloat[\label{fig:04d}]{\includegraphics[width=0.32\textwidth]{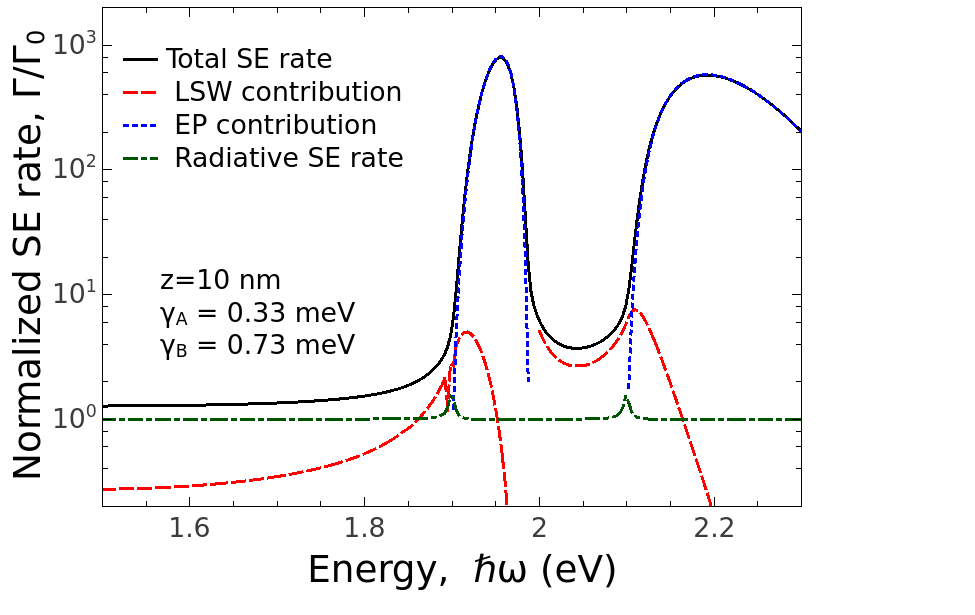}

}

\caption{(Color online) The total normalized spontaneous emission of a QE,
$\tilde{\Gamma}$, placed at a fixed position as a function of its
emission energy, is analyzed with respect to the lossy-surface-wave,
surface-mode and radiative emission contributions. The transition
dipole moment of the QE is oriented along $x$ (a-b) and $z$ (c-d).
(a,c) $\mathbf{r}_{\text{QE}}=(0,0,2\,\text{nm})$. (b,d) $\mathbf{r}_{\text{QE}}=(0,0,10\,\text{nm})$.
The values of the damping parameters considered are $\gamma_{A}=0.3\,\text{meV}$
and $\gamma_{B}=0.7\,\text{meV}$.\label{fig:04}}
\end{figure*}
In this section we will investigate the interaction between a QE and
a MoS\textsubscript{2} layer. In Fig.~\ref{fig:04} we investigate
the spontaneous emission rate when we consider a QE at a fixed position
above a free-standing MoS\textsubscript{2} layer, as a function of
its emission energy, and both $x$- and $z$- orientations for the
transition dipole moment of the QE are considered, (\ref{fig:04a}-\ref{fig:04b})
and (\ref{fig:04c}-\ref{fig:04d}) respectively. The total spontaneous
emission is analyzed over the lossy-surface-wave, exciton polariton
mode and radiative emission contributions. The damping parameters
have the values $\hbar\gamma_{A}=0.33\,\text{meV}$ and $\hbar\gamma_{B}=0.70\,\text{meV}$.

In Fig.~\ref{fig:04a} the QE is positioned at $\mathbf{r}_{\text{QE}}=(0,0,2\,\text{nm})$.
Due to the orientation of the transition dipole moment along $x$,
TE and TM exciton polariton modes are supported by a MoS\textsubscript{2}
layer, depending on the sign of the $\text{Im}\left(\sigma_{\text{res}}\right)$,
as we have already discussed in Sec.~\ref{sec:Sec.IIb}. The contributions
of these modes are obtained by extracting the pole contributions from
Eq.~\eqref{eq:05b}, which for $\varepsilon_{1}=\varepsilon_{2}=1$,
have the form:\begin{widetext}
\begin{equation}
\tilde{\Gamma}_{x,\text{EP}}\left(\omega,\mathbf{r}_{\text{QE}}\right)=\frac{3\pi c}{4\omega}\text{Im}\left[\frac{\alpha^{2}k_{0}^{2}}{k_{z1}^{\text{TE}}}\,\text{e}^{2ik_{z1}^{\text{TE}}z_{\text{QE}}}\Theta\left(-\text{Im}\left(\sigma_{\text{res}}\right)\right)-\frac{k_{z1}^{\text{TM}}}{\alpha^{2}}\,\text{e}^{2ik_{z1}^{\text{TM}}z_{\text{QE}}}\Theta\left(\text{Im}\left(\sigma_{\text{res}}\right)\right)\right]\label{eq:11}
\end{equation}
where $k_{z1}^{i}=\sqrt{k_{0}^{2}-\left(k_{\text{EP}}^{i}\right)^{2}}$,
for $i=\text{TE},\text{TM}$ and where $k_{\text{EP}}^{i}$ are given
by Eqs.~\eqref{eq:09} and \eqref{eq:10}, respectively. The LSWs
contribution is obtained in the large $k_{s}$ limit of the integrand
of Eq.~\eqref{eq:05b} \cite{Ford1984}and has the form
\begin{equation}
\tilde{\Gamma}_{x,\text{LSW}}\left(\omega,\mathbf{r}_{\text{QE}}\right)=\frac{3c}{4\omega}\text{Im}\left[\int\limits _{K}^{\infty}\text{d}k_{s}\left(\frac{-\alpha k_{0}}{ik_{s}+\alpha k_{0}}+\frac{1}{k_{0}^{2}}\frac{i\alpha k_{s}^{3}}{k_{0}+i\alpha k_{s}}\right)\text{e}^{-2k_{s}z_{\text{QE}}}\right],\label{eq:12}
\end{equation}
where the lower limit on the integral is used for numerical reasons
to separate the various contributions to the full integral. In particular,
when there are no TM exciton polariton modes and the TE exciton polariton
modes lie very close to the light line, $K\simeq k_{0}$. When the
TM modes are present, the lower integration limit should be $K>\text{Re}\left(k_{\text{EP}}^{\text{TM}}\right)$,
in order not to include the pole contribution, given by Eq.~\eqref{eq:11}.
The LSWs are non-propagating dissipative modes. The radiative contribution
is given by integrating Eq.~\eqref{eq:05b} over the interval $[0,k_{0}].$
\end{widetext}

In Fig.~\ref{fig:04a}, for emission energies of the QE below the
first exciton energy, $\hbar\omega<E_{A}$, the QE's near-field can
excite LSWs and these lossy modes are the main contribution to the
total SE rate of the QE. Exciting the TE exciton polariton mode makes
a small contribution to the total SE rate. As the emission energy
of the QE is increased, in the interval $E_{A}<\hbar\omega<2\,\text{eV}$,
the TM exciton polariton mode contribution dominates as the main channel
of relaxation for the total SE rate, although the LSW still have a
considerable contribution. At emission energies in the interval $2\,\text{eV}<\hbar\omega<E_{B}$,
the LSW again dominate and the contribution of the TE exciton polariton
modes is small. As we have already argued, the TE exciton polariton
modes are loosely confined to the MoS\textsubscript{2} layer, and
thus their contribution to the normalized SE rate is small, see Eq.~\eqref{eq:11}.
Finally, for emission energies $\hbar\omega>E_{B}$, the TM exciton
polariton mode contribution dominates and the LSW is suppressed, although
its contribution is still considerable. 

In Fig.~\ref{fig:04b} we observe that the enhancement of the total
normalized SE rate of the QE placed at $\mathbf{r}_{\text{QE}}=(0,0,10\,\text{nm})$,
is smaller when compared with the case presented in Fig.~\ref{fig:04a}.
This is due to the fact that the near-field of the QE decouples from
the MoS\textsubscript{2} layer as the QE\textendash MoS\textsubscript{2}
layer distance is increased. Thus, the LSW contribution to the total
SE rate along the whole spectrum is small. The LSWs can only be excited
at small QE\textendash MoS\textsubscript{2} separations. The TE modes
also have a small contribution to the total SE rate, but they can
now compete with the LSWs. However, the TM exciton polariton modes,
in the interval where they are excited, dominate the total SE rate
of the QE. The SE rate is enhanced several orders of magnitude in
those intervals, compared with the free-space value.

When the transition dipole moment of the QE is along $z$, the pole
contribution to Eq.~\eqref{eq:05a} comes exclusively from the TM
exciton polariton mode and has the form:
\begin{equation}
\tilde{\Gamma}_{z,\text{EP}}\left(\omega,\mathbf{r}_{\text{QE}}\right)=-\frac{3\pi c^{2}}{2\omega^{2}}\text{Im}\left[\frac{\left(k_{\text{EP}}^{\text{TM}}\right)^{2}}{\alpha}\,\text{e}^{2ik_{z1}^{\text{TM}}z_{\text{QE}}}\right],\label{eq:13}
\end{equation}
where $k_{z1}^{\text{TM}}=\sqrt{k_{0}^{2}-\left(k_{\text{EP}}^{\text{TM}}\right)^{2}}$.
The LSW contribution, obtained in the limit $k_{s}\to\infty$ of Eq.~\eqref{eq:05a},
has the form
\begin{equation}
\tilde{\Gamma}_{z,\text{LSW}}\left(\omega,\mathbf{r}_{\text{QE}}\right)=\frac{3c}{2\omega k_{0}^{2}}\text{Im}\left[\int\limits _{K}^{\infty}\text{d}k_{s}\frac{i\alpha k_{s}^{3}}{k_{0}+i\alpha k_{s}}\text{e}^{-2k_{s}z_{\text{QE}}}\right],\label{eq:14}
\end{equation}
where the lower limit is determined by the existence of a TM exciton
polariton mode, $K>k_{\text{EP}}^{\text{TM}}$ when present, and by
$K\geqslant k_{0}$ when absent. The radiative contribution, $\tilde{\Gamma}_{z,0}\left(\omega,\mathbf{r}_{\text{QE}}\right)$,
is given by integrating Eq.~\eqref{eq:05a} over the interval $\left[0,k_{0}\right]$.

In Fig.~\ref{fig:04c} we investigate the case for which the transition
dipole moment of the QE is along $z$, at $\mathbf{r}_{\text{QE}}=(0,0,2\,\text{nm})$,
showing the SE rate as a function of the QE emission energy. Due to
the dipole orientation, only TM exciton polariton modes are excited
in the intervals $E_{A}<\hbar\omega<2\,\text{eV}$ and $\hbar\omega>E_{B}$
and these are the main channels of relaxation of the QE. The LSWs
dominate the total SE rate outside the interval where TM exciton polariton
modes are excited, although they also make a considerable contribution
in the range $E_{A}<\hbar\omega<2\,\text{eV}$. In Fig.~\ref{fig:04d}
the distance between the QE and the MoS\textsubscript{2} layer is
increased to $\mathbf{r}_{\text{QE}}=(0,0,10\,\text{nm})$. The LSWs
contribution decreases as the distance between QE and the MoS\textsubscript{2}
layer increases. The total SE rate is enhanced several orders of magnitude
when the TM exciton polariton mode is excited. In general, the SE
rate of a QE has similar characteristics for the $x$ and $z$ polarizations.
The main difference is that, for a QE with a transition dipole moment
along $x$, TE exciton polariton modes can be excited. The coupling
between a QE and the MoS\textsubscript{2} layer is more efficient
for the $z$ orientation. For the rest of this paper we focus on a
QE with $z$ orientation of the transition dipole moment. 

The distance dependence of the interaction between QEs and a TMD monolayer
has been investigated experimentally and different theoretical expressions
have been used to fit the experimental results. In ref.~\cite{Goodfellow2016}
they report a $z^{-4}$ behavior of the distance dependence of the
interaction between a QE and a MoSe\textsubscript{2} layer, although
the authors fit the intensity quenching rather than the lifetime quenching.
On the other hand, ref.~\cite{Sampat2016} uses multiple QEs and
investigates their lifetime quenching in the presence of a MoS\textsubscript{2}
monolayer. The extracted fittings for the lifetime quenching are between
$z^{-3}$ to $z^{-4}$.

\begin{figure*}
\subfloat[\label{fig:05a}]{\includegraphics[width=0.35\textwidth]{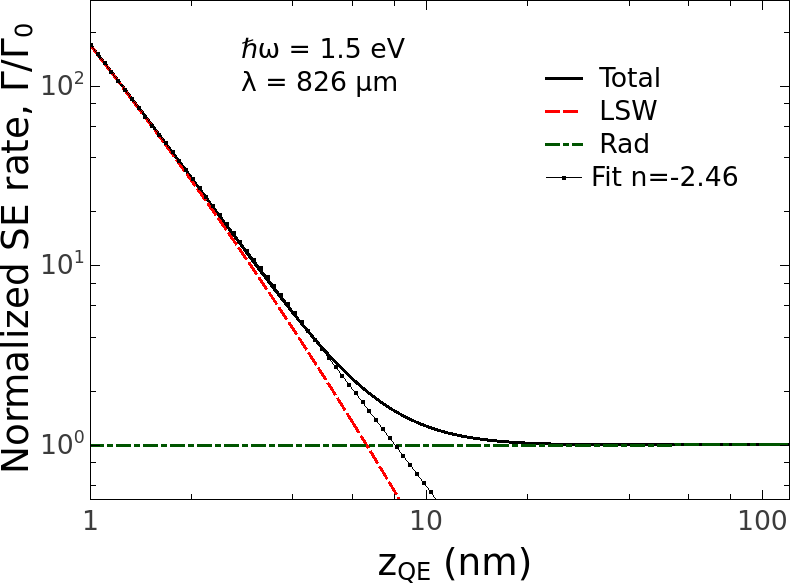}

}~~~\subfloat[\label{fig:05b}]{\includegraphics[width=0.35\textwidth]{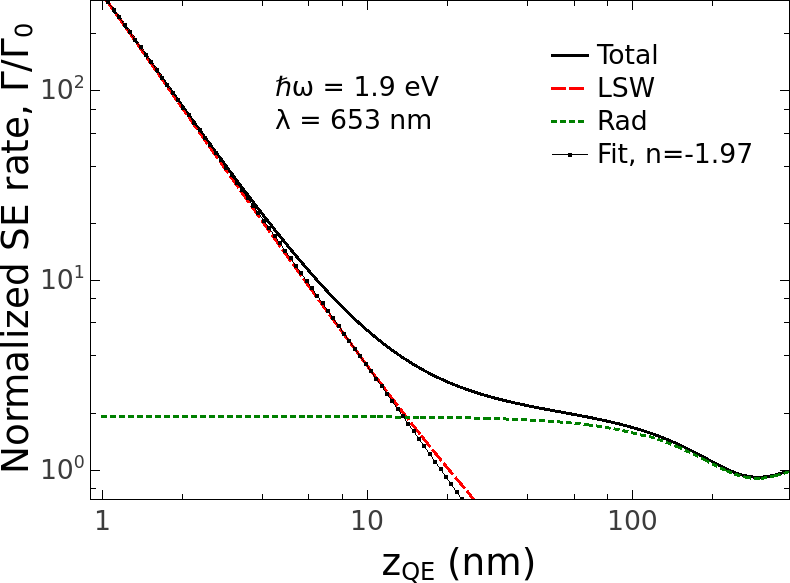}

}

\subfloat[\label{fig:05c}]{\includegraphics[width=0.35\textwidth]{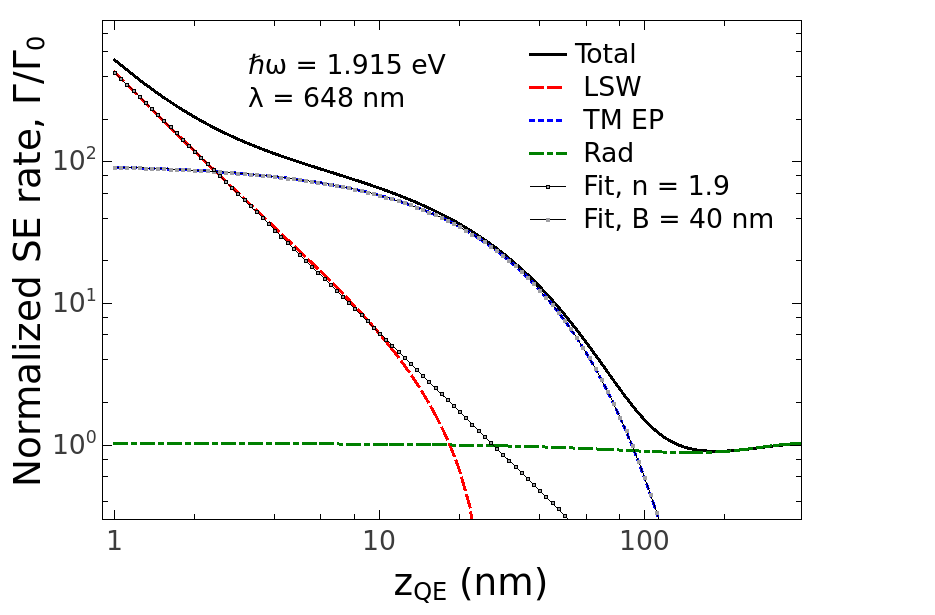}

}~~~\subfloat[\label{fig:05d}]{\includegraphics[width=0.35\textwidth]{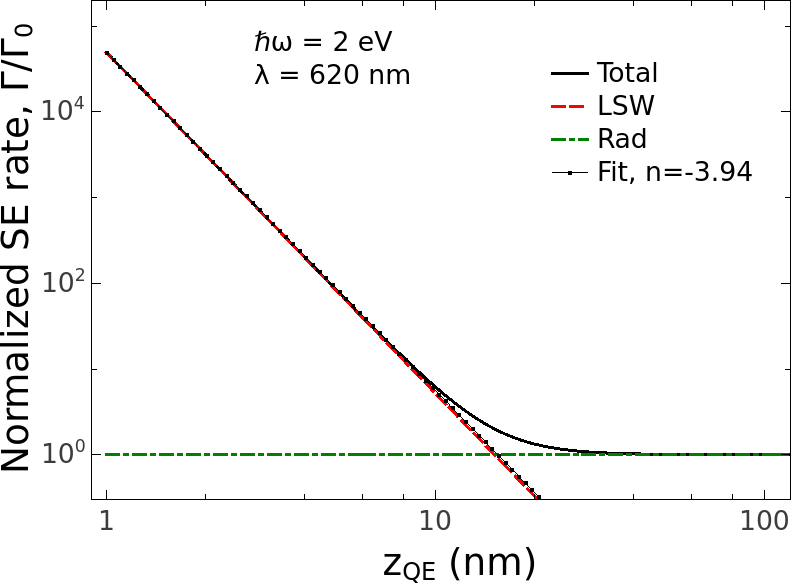}

}

\caption{(Color online) The total normalized spontaneous emission rate of a
QE, with fixed emission energy as a function of its position $\mathbf{r}=(0,0,z_{\text{QE}})$,
is analyzed with respect to the lossy-surface-wave, exciton polariton
mode and radiative emission contributions. The transition dipole moment
of the QE is $z$-oriented. (a) $\hbar\omega=1.5\,\text{eV}$. (b)
$\hbar\omega=1.9\,\text{eV}$. (c) $\hbar\omega=1.915\,\text{eV}$
(d) $\hbar\omega=2.0\,\text{eV}$. The damping parameters have the
values, $\gamma_{B}=0.3\,\text{meV}$ and $\gamma_{A}=0.7\,\text{meV}$.\label{fig:05}}
\end{figure*}
 In Fig.~\ref{fig:05} we present the distance dependence of the
spontaneous emission rate of a QE, placed at $\mathbf{r}_{\text{QE}}=(0,0,z_{\text{QE}})$,
and oriented along $z$, for fixed emission energies. We analyze the
different contributions to the total SE rate, the LSWs, TM exciton
polariton modes and the radiative emission using Eqs.~\eqref{eq:05a},
\eqref{eq:13} and \eqref{eq:14}. We consider four values of the
emission energy of the QE, one in the range $\hbar\omega<E_{A}$,
where the LSWs dominate and one in the $E_{A}<\hbar\omega<2\,\text{eV}$
range, where the TM exciton polariton modes are excited. The other
two values are at $\hbar\omega=1.9\,\text{eV}$, right on the exciton
energy $E_{A}$, and $\hbar\omega=2.0\,\text{eV}$, at the position
where $\text{Im}\left(\sigma_{\text{res}}\right)$ changes sign due
to the interaction between the two excitons. 

In Fig.~\ref{fig:05a} the QE emission energy is $\hbar\omega=1.5\,\text{eV}$,
in the interval $\hbar\omega<E_{A}$, thus we see that the main contribution
comes from the LSWs very close to the MoS\textsubscript{2} layer,
but this channel of interaction dies out quickly and, at separations
as small as $z_{\text{QE}}\approx8\,\text{nm}$, the SE rate reverts
to its free-space value. At this energy there is no exciton polariton
mode, due to the dipole moment orientation of the QE. The integral
in Eq.~\eqref{eq:14} has contributions of the form $A_{1}/z^{2}+A_{2}/z^{3}+A_{3}/z^{4}$,
therefore in order to analyze the LSW contribution, we use the fitting
expression:
\begin{equation}
f(z)=A\,z^{n}\label{eq:15}
\end{equation}
and in Fig.~\ref{fig:05a} we show that $n=-2.5$. This fitting shows
that the behavior of MoS\textsubscript{2} layer is very different
to the case of a graphene layer in the optical part of the spectrum.
The optical response of graphene, in the optical part of the spectrum,
is constant and characterized by a surface conductivity of $\sigma_{\text{Graph}}=\sigma_{0}=e^{2}/2\hbar$.
The distance dependence of the SE rate of a QE is then given by $\tilde{\Gamma}\propto1/z^{4}$,
Eq.~\eqref{eq:14}, which is a universal scaling law of the distance
dependence between a QE and a graphene monolayer, in the optical part
of the spectrum \cite{Gaudreau2013,Lee2014a,Chen}. In Fig.~\ref{fig:05b}
the QE energy is $\hbar\omega=E_{A}=1.9\,\text{eV}$, and we observe
a behavior similar to Fig.~\ref{fig:05a}, but now the fitting of
the LSWs, which have the largest contribution to the total SE rate,
gives $n\approx-2$ and the QE reverts to the radiative value of the
SE at distances of $10\,\text{nm}$. We furthermore observe that the
radiative SE rate of the QE is enhanced very close to the MoS\textsubscript{2},
which is an effect of constructive interference with the image dipole,
due to the dipole orientation.

In Fig\@.~\ref{fig:05c} the emission energy of the QE is $\hbar\omega=1.915\,\text{eV}$
and we observe that the main contribution to the SE rate close to
the MoS\textsubscript{2} layer again comes from the LSWs. At this
energy a TM exciton polariton mode is excited and thus adds a new
path of relaxation for the QE. The TM exciton polariton mode has a
considerable contribution at small separations between the QE and
the MoS\textsubscript{2} layer and dominates at intermediate distances,
$6\,\text{nm}<z_{\text{QE}}<100\,\text{nm}$. In order to better understand
the influence of the TM modes on the SE rate, we use a fitting expression
of the form:
\begin{equation}
g(z_{\text{QE}})=A\,\exp(-2z_{\text{QE}}/B),\label{eq:16}
\end{equation}
where $B$ is the fitting parameter of interest, connected with the
penetration depth of the TM exciton polariton mode, $\delta_{\text{EP}}=1/\text{Im}\left(k_{z}^{\text{EP}}\right)$.
The value of the fitting parameter in Fig.~\ref{fig:05c} is found
to be $B=40\,\text{nm}$ which is the same as the value plotted in
Fig.~\ref{fig:03c} where $\delta_{\text{TM}}=40\,\text{nm}$. The
distance dependence of LSWs is described by Eq.~\eqref{eq:15} with
$n\approx-2$. The LSW contribution to the SE rate is calculated using
the approximate expression Eq.~\eqref{eq:14}. When the TM modes
are also present, it becomes more challenging to distinguish between
the propagating, TM exciton polariton, and non-propagating, LSW, nature
of the relaxation. In Fig.~\ref{fig:05d} the emission energy is
$\hbar\omega=2\,\text{eV}$, and at this energy there is a change
of sign of the $\text{Im}(\sigma_{\text{MoS}_{2}})$ from positive
to negative values while the ratio $\text{Re}\left(\sigma_{\text{res}}\right))/\text{Im}\left(\sigma_{\text{res}}\right)\gg1$,
thus we can safely ignore the imaginary part. We then have a situation
identical to graphene and the LSWs, which dominate at small QE\textendash MoS\textsubscript{2}
layer separations, follow a behavior given by Eq.~\eqref{eq:15}
with $n\approx-4$.

\begin{figure}
\includegraphics[width=0.45\textwidth]{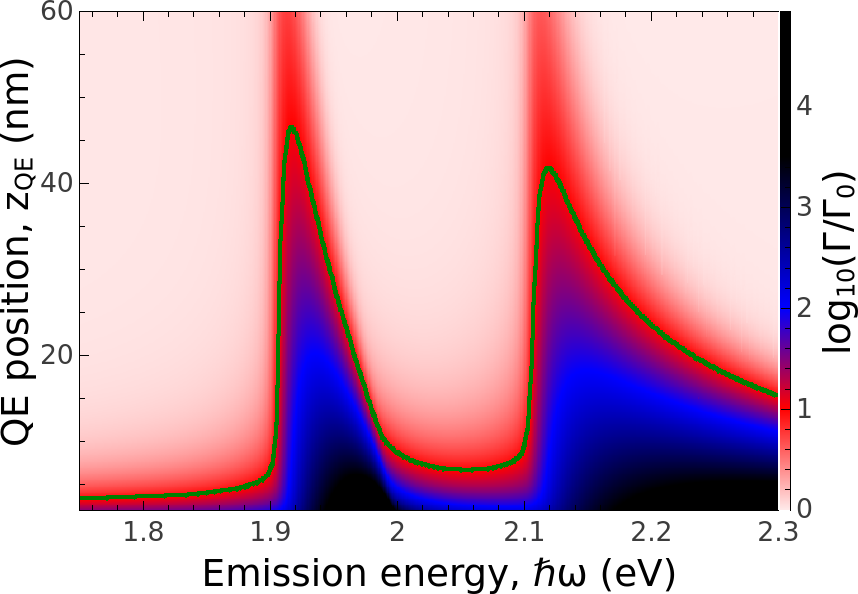}\caption{(Color online) Contour plot of the total normalized SE of a QE, as
a function of its position, $\mathbf{r}_{\text{QE}}=(0,0,z_{\text{QE}})$,
and emission energy, $\hbar\omega$. The transition dipole moment
of the QE is $z$-oriented. The value of the damping parameters that
are considered are $\gamma_{A}=0.3\,\text{meV}$ and $\gamma_{B}=0.7\,\text{meV}$.\label{fig:06}}
\end{figure}
For completeness we present the full spectral and distance dependence
of the SE rate of a QE in the presence of a MoS\textsubscript{2}
layer, in Fig.~\ref{fig:06}. This is a contour plot of the normalized
total SE rate of a QE, as a function of the QE position, $\mathbf{r}_{\text{QE}}=(0,0,z_{\text{QE}})$,
and its emission energy, $\hbar\omega$. The transition dipole moment
of the QE is along $z$. The olive green line represents the boundary
of the parameter space where $\Gamma/\Gamma_{0}>10$. We observe that
at emission energies where one can excite the TM exciton polariton
mode supported by the MoS\textsubscript{2} layer, at $1.9\,\text{eV}<\hbar\omega<2\,\text{eV}$
and $\hbar\omega>2.1\,\text{eV}$, the SE rate is enhanced up to $10$
times, compared with its free-space value, for distances up to $40\,\text{nm}$.
At small distances, the SE rate is enhanced due to the excitation
of the non-propagating LSWs. The values of the damping parameters
considered are $\hbar\gamma_{A}=0.3\,\text{meV}$ and $\hbar\gamma_{B}=0.7\,\text{meV}$. 

\subsection{Spontaneous emission in the presence of a superlattice of MoS\protect\textsubscript{2}
layers\label{sec:Sec.IIIC}}

In this section we investigate the influence of the presence of a
superlattice composed of multiple MoS\textsubscript{2} layers on
the emission properties of a QE. There are contradicting experimental
reports regarding the influence on the SE rate of a QE interacting
with TMD layers, as the number of layers is increased. In particular,
in refs.~\cite{Prins2014,Chen}, the authors report that, as the
number of MoS\textsubscript{2} layers is increased, the SE rate of
the QEs decreases. The authors of ref.~\cite{Chen} use a bulk dielectric
permittivity to describe the optical response of the MoS$_{2}$ and
they attribute the decreasing behavior to dielectric screening \cite{Gordon2013}.
In particular, they found that, by increasing the thickness of the
MoS\textsubscript{2} slab, the field intensity created by a dipole
source on the slab drops. The screening effect is connected with the
difference between the parallel and perpendicular dielectric permittivities
of the MoS\textsubscript{2} slab, more details can be found in ref.~\cite{Chen}.
Also, the real part of the dielectric permittivity has larger values
compare with the imaginary part, further increasing the screening
effect \cite{Gordon2013}. Their analysis is focused on a single emission
energy of the QE. On the other hand, in ref.~\cite{Zang2016}, the
authors report an opposite behavior where, as the number of layers
of SnS\textsubscript{2} is increased, the SE rate also increases.
This discrepancy is attributed to the fact that the MoS\textsubscript{2}
material exhibits a band inversion from indirect, as a bulk material,
to direct as a monolayer, while SnS\textsubscript{2} is an indirect
band gap material down to a monolayer. Furthermore, in refs.~\cite{Prins2014,Chen,Zang2016},
the emission profile of the QEs investigated is different for each
case.

Our analysis follows a different path. Instead of using a slab for
approximating the MoS\textsubscript{2} layer, and describing its
optical response through an anisotropic dielectric permittivity, we
treat the MoS\textsubscript{2} as a 2D material, whose optical response
is given by Eq.~\eqref{eq:07}. We describe the interaction between
a QE and a MoS\textsubscript{2} superlattice using Eqs.~\eqref{eq:A02}-\eqref{eq:A03}.
Multiple scattering between the MoS\textsubscript{2} layers of the
MoS\textsubscript{2} creates a number of modes, depending on the
number of layers. We analyze and investigate the influence these mode
have on the total SE rate of the QE. We choose to investigate free
standing MoS\textsubscript{2} superlattices for simplicity. The inclusion
of the substrate will slow down the SE rate due to the difference
between the substrate and superlattice dielectric permittivities.
The main relaxation path for a QE is associated with the exciton polariton
modes, provided by the MoS\textsubscript{2} superlattice. Their existence
is unaffected by the inclusion of a substrate.

\begin{figure*}
\subfloat[\label{fig:07a}]{\includegraphics[width=0.32\textwidth]{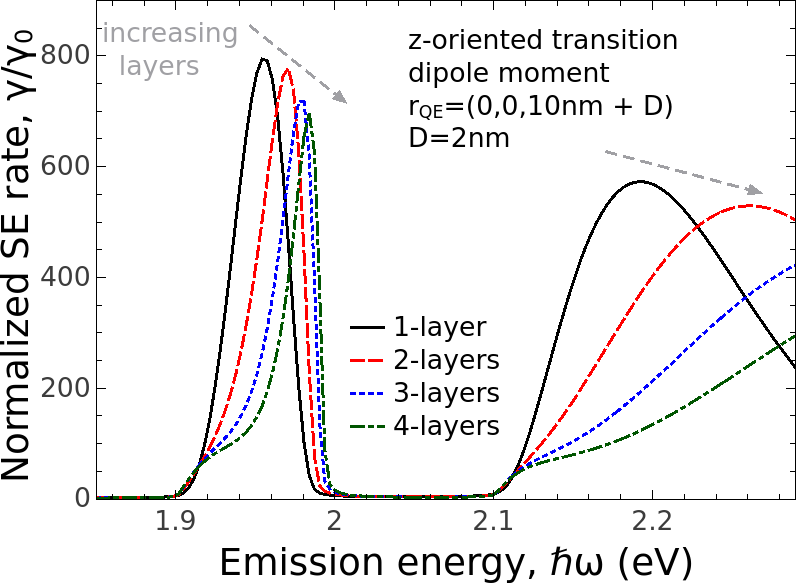}}~~~\subfloat[\label{fig:07b}]{\includegraphics[width=0.32\textwidth]{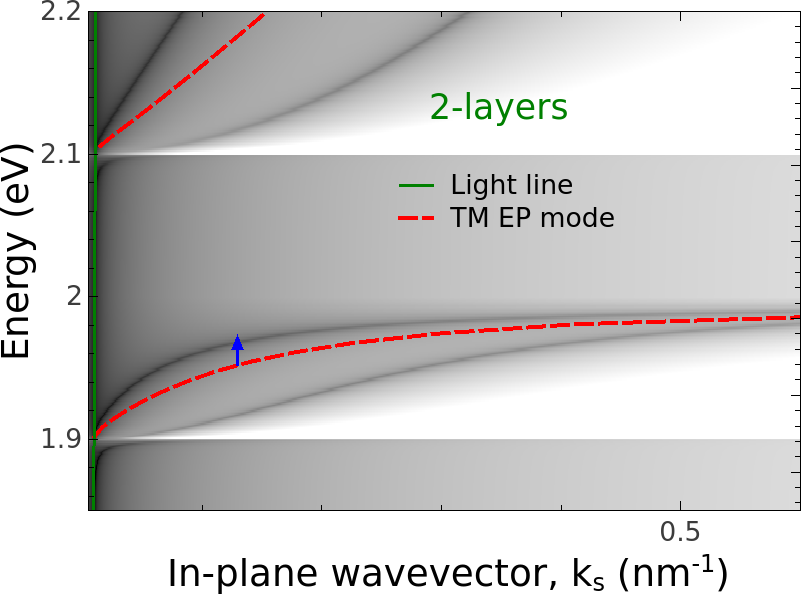}

}~~~\subfloat[\label{fig:07c}]{\includegraphics[width=0.32\textwidth]{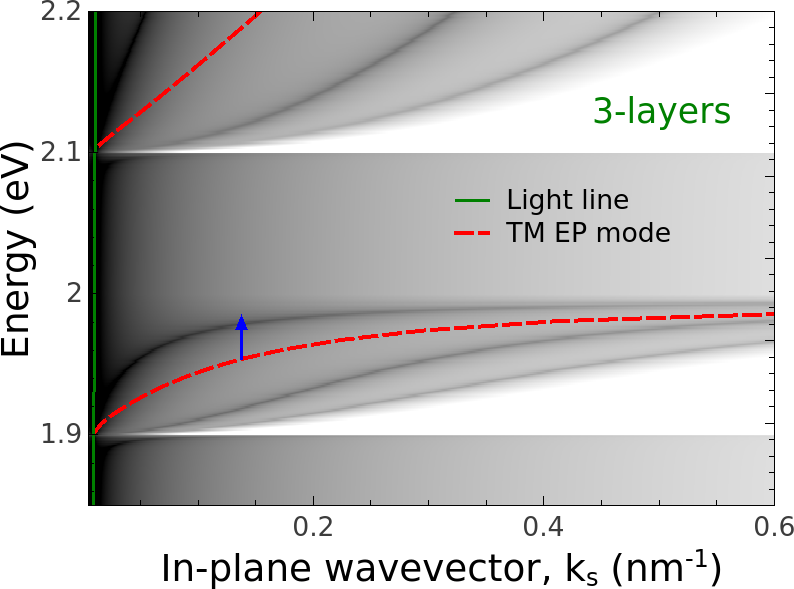}

}\caption{(Color online) (a) Total normalized spontaneous emission of a QE,
placed at a fixed position, $\mathbf{r}_{\text{QE}}=(0,0,10\,\text{nm}+D)$,
as a function of its emission energy, considering different numbers
of MoS\protect\textsubscript{2} layers. The transition dipole moment
of the QE is $z$-oriented. (b-c) Dispersion relation, $k_{s}(\omega)$,
showing $\log\left(\left|R_{N}\right|\right)$, for $\text{Im}(\sigma_{\text{res}})>0$,
and $\log\left(\left|R_{M}\right|\right)$, for $\text{Im}(\sigma_{\text{res}})<0$,
considering multiple MoS\protect\textsubscript{2} layers. (b) 2-layers.
(d) 3-layers. $D$ is the thickness between the lower and upper MoS\protect\textsubscript{2}
layers, here $D=2\,\text{nm}$. The dispersion relation of the TM
exciton polariton mode of a single MoS\protect\textsubscript{2} layer
is presented with a red dashed line line. The damping parameters have
the values, $\gamma_{B}=0.3\,\text{meV}$ and $\gamma_{A}=0.7\,\text{meV}$.
\label{fig:07}}
\end{figure*}
In Fig.~\ref{fig:07a} we present the SE rate as a function of emission
energy considering the interaction between a QE and a monolayer and,
double, triple and quaduple MoS\textsubscript{2} layers. Again we
consider the case where the transition dipole moment of the QE is
oriented along $z$. The position of the QE is fixed at $\mathbf{r}_{\text{QE}}=(0,0,10\,\text{nm}+D)$,
where $D$ is the distance between the top and bottom layers. We consider
a fixed value of this thickness, $D=2\,\text{nm}$, and the distance
between the layers is kept equal. So as the number of layers is increased
the distance between them is decreased. In Fig.~\ref{fig:07a} we
observe that, as the number of layers increases, the peak value of
the normalized SE rate blue-shifts and the absolute value of its enhancement
decreases. The shift is smaller when the lower TM exciton polariton
mode is excited. In order to give an explanation for this effect we
present in Figs.~(\ref{fig:07b},\ref{fig:07c}) the dispersion relation
for two superlattice examples.

As we have already discussed, the TE and TM exciton polariton modes
are obtained as poles of the generalized Fresnel reflection coefficients.
For a superlattice nanostructure more details are given in App.~\ref{sec:AppA}.
In Figs.~(\ref{fig:07b},\ref{fig:07c}) we present a contour plot
of the logarithm of the absolute value of the reflection coefficients
$R_{N}(k_{s},\omega)$, for $E_{A}<\hbar\omega<2.0\,\text{eV}$ and
$\hbar\omega>E_{B}$, and $R_{M}(k_{s},\omega)$, for $\hbar\omega<E_{A}$
and $2.0\,\text{eV}<\hbar\omega<E_{B}$, as a function of the in-plane
wave vector, $k_{s}$, and the energy, $\hbar\omega$. The generalized
reflection coefficients, $R$, are calculated by solving Eq.~\eqref{eq:A04}.
The TM mode has the largest contribution to the SE rate, see Fig.~\ref{fig:07a}.
The dispersion relation lines are given by the dark color lines in
the contour plot. We observe that as the number of layers increases,
more branches emerge in the energy range where TM modes are supported
by the MoS\textsubscript{2} superlattice, and the number of branches
is equal the number of layers. These extra branches are connected
with the multiple scatterings in the MoS\textsubscript{2} superlattice.
In the same figure we present with a red dashed line the dispersion
relation of a single MoS\textsubscript{2} layer for direct comparison.

The peak in the SE rate enhancement of a QE for a single layer is
at $\hbar\omega=1.95\,\text{eV}$, Fig.~\ref{fig:07a}. The main
channel of relaxation of the QE, in the presence of the MoS\textsubscript{2}
superlattice, is the TM exciton polariton mode. We choose to focus
on the lower branch of the TM exciton polariton modes, related to
the first peak of the normalized SE rate in Fig.~\ref{fig:07a}.
The peak value for the single layer is connected with the penetration
depth, $\delta_{\text{EP}}^{\text{TM}}=1/\text{Im}\left(k_{z}^{\text{TM}}\right)$,
where $k_{z}^{\text{TM}}=\sqrt{k_{0}^{2}-k_{\text{EP}}^{\text{TM}}}\approx ik_{\text{EP}}^{\text{TM}}$
and since $k_{\text{EP}}^{\text{TM}}\gg k_{0}$, we find $\delta_{\text{EP}}^{\text{TM}}=1/\text{Re}\left(k_{\text{EP}}^{\text{TM}}\right)$.
In Eq.~\eqref{eq:13}, the exciton polariton contribution to the
SE rate depends on a factor $c^{2}/\omega^{2}$, thus for the same
value of $k_{\text{EP}}^{\text{TM}}$, there is a decrease in the
absolute value of the normalized SE rate with increasing energy, explaining
the trend we observe in Fig.~\ref{fig:07a}. In Figs.~\ref{fig:07b}
and \ref{fig:07c}, we show the position of the peak value of the
SE rate of the QE interacting with a single MoS\textsubscript{2}
layer and the blue arrow indicates the blue shift of the energy at
which the peak value of the SE rate emerges in Fig.~\ref{fig:07a}
when the double layer is considered. For the single layer, the peak
of the SE rate is at $\hbar\omega=1.95\,\text{eV}$ at $k_{\text{EP}}^{\text{TM}}=0.1\,\text{nm}^{-1}$,
$\delta_{\text{EP}}^{\text{TM}}=10\,\text{nm}$, while for the double
layer the peak is at $\hbar\omega=1.97\,\text{nm}$ and for the triple
layer it is at $\hbar\omega=1.98\,\text{nm}$. Thus the dispersion
relations give an explanation for the blue shift of the peak value
of the normalized SE rate.

Therefore, to investigate the interaction between a specific QE and
a MoS\textsubscript{2} supelattice, one must take into account the
reduced interaction between QEs\textendash MoS\textsubscript{2} superlattice
as one starts increasing the number of layers. While the emission
properties of the QE do not change, its environment is modified as
there is a redistribution of the available modes. The dispersion relation
plays a crucial role in explaining this effect, giving us the available
modes that can be supported. 

\begin{figure}
\subfloat[\label{fig:08a}]{\includegraphics[width=0.45\textwidth]{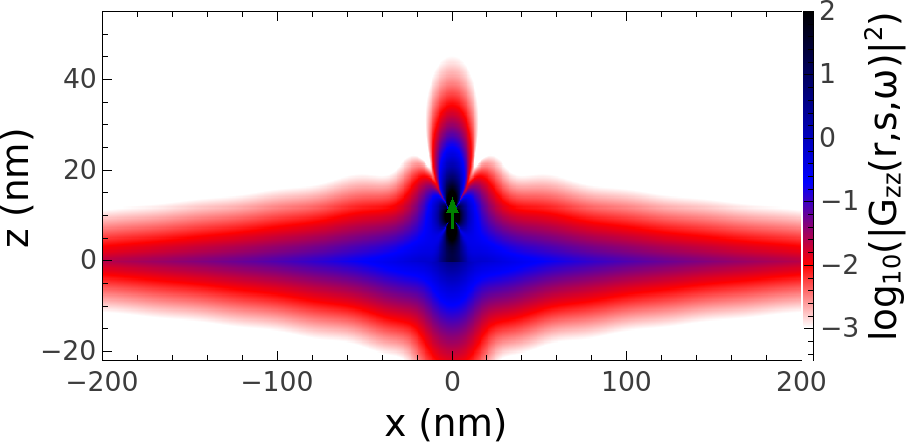}

}

\subfloat[\label{fig:08b}]{\includegraphics[width=0.45\textwidth]{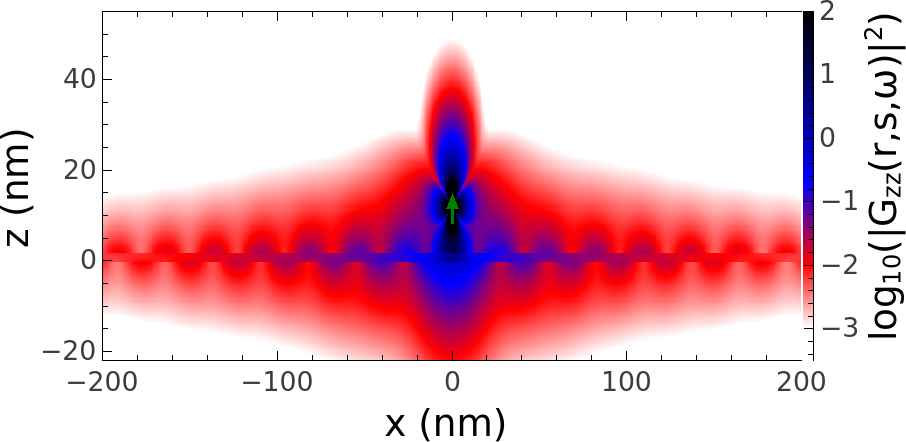}

}

\subfloat[\label{fig:08c}]{\includegraphics[width=0.45\textwidth]{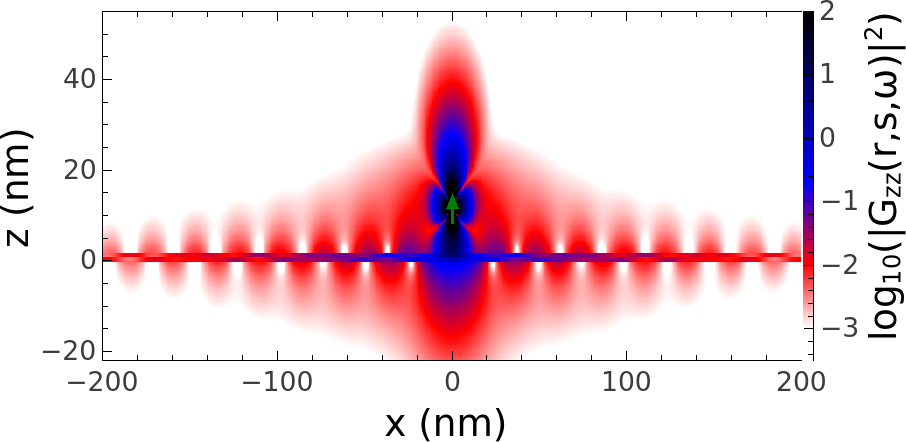}

}\caption{(Color online) Contour plots of the logarithm of the field intensity,
$\log_{10}\left(\left|G_{zz}(\mathbf{r},\mathbf{s},\omega)\right|^{2}\right)$,
created by a QE, placed at $\mathbf{s}=(0,0,D+10\,\text{nm})$. The
transition dipole moment of the QE is along $z$ and its emission
energy is $\hbar\omega=1.915\,\text{eV}$. (a) Single layer, (b) Double
layer, (c) Triple layer. The damping parameters have the values, $\gamma_{B}=0.3\,\text{meV}$
and $\gamma_{A}=0.7\,\text{meV}$. \label{fig:08}}
\end{figure}
To further analyze this effect, in Fig\@.~\ref{fig:08} we present
contour plots of the logarithm of the absolute value of the electric
field, $E_{z}(\mathbf{r},\omega)\propto\mathfrak{G}_{zz}(\mathbf{r},\mathbf{r}_{\text{QE}},\omega)$,
created by a QE placed at $\mathbf{r}_{\text{QE}}=(0,0,10\,\text{nm}+D)$,
in the presence of a MoS\textsubscript{2} superlattice, $D=2\,\text{nm}$
in our case. The emission energy of the QE has been selected to be
at the maximum value of the SE rate for a single MoS\textsubscript{2}
layer, $\hbar\omega=1.95\,\text{eV}$. The scale of the color maps
in Fig.~\ref{fig:08} is the same in all panels, for direct comparison.
We observe that the field intensity decreases as the number of the
MoS\textsubscript{2} layers increases. This is due to the poorer
coupling of the near field of the QE to the MoS\textsubscript{2}
superlatice. We observe also that the extent of the field in the $x$
direction decreases as the number of layers is increased. The propagation
length for the single layer is $L_{\text{EP}}^{\text{TM}}=373\,\text{nm}$
while for the double layer it is $L_{\text{EP}}^{\text{TM}}=300\,\text{nm}$,
and, as the number of layers further increases, the propagation length
further reduces.

The opposite behavior can be observed when the QE emission energy
is at the maximum value of the SE rate for a MoS\textsubscript{2}
superlattice, e.g.~for the three-layer superlattice. Then the SE
rate of the QE decreases with decreasing the number of layers. For
this example of the trilayer, at the resonance $\hbar\omega=1.98$,
the normalized SE rate is $\tilde{\Gamma}=718$, compared with $\tilde{\Gamma}=444$
for the two layer and $\tilde{\Gamma}=102$ for a monolayer. 

\section{Conclusions and future work\label{sec:IV}}

In this contribution we have investigated the spectral and distance
dependence of the SE rate of a QE in the presence of a MoS\textsubscript{2}
layer and superlattice. A MoS\textsubscript{2} layer supports transverse
electric and transverse magnetic surface exciton polariton modes.
The TM modes are strongly confined to the MoS\textsubscript{2} layer
and have long propagation lengths. The TE modes are only loosely confined
to the MoS\textsubscript{2} layer.

The total SE rate of the QE in the presence of a MoS\textsubscript{2}
layer is analyzed with respect to the different contributions, namely,
the lossy surface wave, surface exciton polariton and radiative emission.
In the main part of the discussion we showed that the existence of
TE and TM exciton polariton modes is connected with the surface conductivity
of the MoS\textsubscript{2}, specifically with the sign of its imaginary
part. For energies at which the TM exciton polariton modes are excited,
the SE rate of a QE is enhanced several order of magnitude, compared
with its free-space value. For all the emission energies of the QE,
the main contribution to the SE comes from the LSW at small separations,
but their contribution dies out fast as the separation is increased.
When the TM modes are excited, they dominate at intermediate distances,
$6\,\text{nm}$ to $100\,\text{nm}$. For distances at which the LSW
are not excited or above the penetration depth of the TM exciton polariton
modes, the QE radiates to the far field.

Next, the interaction between a QE and a MoS\textsubscript{2} superlattice
is investigated. We observe a blue-shift of the peak value of the
SE rate of the QE, as a function of its emission energy, as we increase
the number of MoS\textsubscript{2} layers from one to four. Using
the dispersion relation plot, this blue-shift is explained. It is
seen that the number of layers determines the number of branches of
the exciton polariton modes available. The blue-shift of the SE rate
is connected with the blue-shift of the dispersion line for the MoS$_{2}$
superlattice, compared with the single layer. For a QE with emission
energy corresponding to the peak energy of the SE rate for the monolayer,
as the number of layers is increased the coupling decreases and the
field intensity distribution around the superlattice decreases. The
opposite behavior is observed if the emission energy of the QE is
on resonance with a MoS\textsubscript{2} superlattice.

Although the results presented in this study focused on MoS\textsubscript{2}
as a material, they are quite general and can be applied to any material
whose optical properties are determined by exciton generation. Thus,
they can be applied to any of the rest of the TMD family. Furthermore,
we have choosen to concentrate on a theoretical investigation of the
MoS\textsubscript{2}, and not to fit existing experimental data.
This is due to the fact that the material parameters are strongly
influenced by the quality of the material sample itself. 

This study made a contribution to explaining all the contradictory
results regarding the spectral and distance dependences of QEs in
the presence of MoS\textsubscript{2} layers. Specifically, we presented
a $z^{n}$, $n=2,3,4$, distance dependence of the SE rate of a QE,
which is connected with the LSWs at the different emission energies.
Also we observed the existence of exciton polariton modes and how
they modify the emission properties of QEs and the strength of the
interaction, where the distance dependence follows the expression
$\sim\exp\left(-2z/\delta_{EP}^{TM}\right)$. Moreover, we explained
that the coupling of a QE with a MoS\textsubscript{2} superlatice
depends on its emission energy. The peak value for the SE rate of
a QE blue-shifts depending on the number of MoS$_{2}$ layers, due
to the different mode distribution supported by these structures.
While the emission properties of the QE remain the same, the modes
provided by the environment change as the number of MoS\textsubscript{2}
layers changes. Depending on the emission energy of the QE, the SE
rate can increase or decrease as the number of MoS$_{2}$ layers is
increased. Dielectric screening can explain certain results for certain
emission energies of the QE.

Multilayer devices based on MoS\textsubscript{2} and graphene can
be the \textcolor{black}{precursors} of an all-optical device. Graphene's
optical properties can be tuned by changing its chemical potential,
but it has small absorption in the visible part of the optical spectrum.
Combining graphene with TMD layers allows one access to the best of
both materials for applications like light harvesting and light emitting
devices \cite{Wang2014,Yi2016}. In general the total absorption of
these two materials can be further enhanced by including also layers
of QEs. The emission properties of the QEs can be chosen depending
on the nature of the application. For light harvesting devices we
need the emission energy of the QEs to maximize the non-radiative
energy transfer to the MoS\textsubscript{2} layer, where the generated
electron-hole pair will be harvested. On the other hand, for LEDs
we need to maximize the far field emission of the SE rate of the QEs.
Both of these effects can be further investigated for devices composed
from patterned 2D nanostructures, like ribbons and disks, where the
redistribution of the available modes gives rise to sharp resonances
\cite{Christensen,Karanikolas2016}.
\begin{acknowledgments}
This work was supported by the Science Foundation Ireland under grant
No. 10/IN.1/12975. 
\end{acknowledgments}

\appendix

\section{Green's tensor for a $\text{MoS}_{2}$ superlattice\label{sec:AppA}}

A multilayer planar geometry is considered, which consists of a number
of $N$ layers, indexed by their layer number $i=1,\ldots,N$ where
each layer has thickness $d_{i}$ and dielectric permittivity $\varepsilon_{i}$.
The layers are of infinite extent in the $xy$ plane and the $z$
axis is perpendicular to the surface of each layer. 

The method of scattering superposition is used \cite{Tai1994,Chew1995}
where the Green's tensor splits into two parts:

\begin{equation}
\mathfrak{G}(\mathbf{r},\mathbf{s},\omega)=\mathfrak{G}_{h}(\mathbf{r},\mathbf{s},\omega)+\mathfrak{G}_{s}(\mathbf{r},\mathbf{s},\omega),\label{eq:A01}
\end{equation}
where $\mathfrak{G}_{h}(\mathbf{r},\mathbf{s},\omega)$ is the homogeneous
part that accounts for direct interaction between the source and target
point at $\mathbf{s}$ and $\mathbf{r}$ respectively, and is non-zero
when both points are in the same media and there is no discontinuity
between them. $\mathfrak{G}_{s}(\mathbf{r},\mathbf{s},\omega)$ is
the scattering part, is always present and accounts for the multiple
reflections and transmissions taking place at the interfaces.

\begin{widetext}The general form of the scattering part of the Green's
tensor has the form:
\begin{equation}
\mathfrak{G}_{s}(\mathbf{r},\mathbf{s},\omega)=\frac{i}{8\pi^{2}}\int\textrm{d}^{2}k_{s}\frac{1}{k_{zi}k_{s}^{2}}\sum_{T}R_{T}^{\pm(ij)\pm}\mathbf{T}(\mathbf{k}_{s},\pm k_{zi},\mathbf{r})\otimes\mathbf{T}^{*}(\mathbf{k}_{s},\pm k_{zj},\mathbf{s}).\label{eq:A02}
\end{equation}

A summation is implied for each pair of $\pm$ indices. These indices
show the direction of propagation of the electromagnetic modes, the
first index for the acceptor and the second for the donor. Also the
summation over $\mathbf{T}$ is over the $\mathbf{M}$ and $\mathbf{N}$
modes which are connected with the transverse electric and transverse
magnetic modes, respectively. The form of $\mathbf{M}$ and $\mathbf{N}$
can be found in ref.~\cite{Chew1995}. For the planar geometries
there are no hybrid modes. The boundary conditions imposed on the
system of multilayers are the continuity condition and the radiation
condition. The first condition is given by continuity equations at
each interface:\begin{subequations}\label{eq:A03}
\begin{equation}
\hat{\mathbf{z}}\times\left.\left[\mathfrak{G}^{(ij)}(\mathbf{r},\mathbf{s},\omega)-\mathfrak{G}^{((i+1)j)}(\mathbf{r},\mathbf{s},\omega)\right]\right|_{z=d_{i}}=0,\label{eq:A03a}
\end{equation}
\begin{equation}
\hat{\mathbf{z}}\times\left.\left[\boldsymbol{\nabla}\times\mathfrak{G}^{(ij)}(\mathbf{r},\mathbf{s},\omega)-\boldsymbol{\nabla}\times\mathfrak{G}^{((i+1)j)}(\mathbf{r},\mathbf{s},\omega)\right]\right|{}_{z=d_{i}}=-i\frac{4\pi}{c}k_{0}\sigma\hat{z}\times\hat{z}\times\mathfrak{G}^{((i+1)j)}(\mathbf{r},\mathbf{s},\omega),\label{eq:A03b}
\end{equation}
\end{subequations}where $\sigma$ is the surface conductivity of
the 2 dimensional material, for our case it is the MoS\textsubscript{2}
layer, Eq.~\eqref{eq:07}.\end{widetext}

By applying these boundary equations, an inhomogeneous system of $2^{N-1}$
equations is defined which have $2^{N-1}$ unknowns, the generalized
$R_{M(N)}^{\pm(ij)\pm}$ coefficients. These coefficients are sufficient
to uniquely determined the problem under consideration through the
exact knowledge of the scattering part of the Green's tensor. In order
to find the generalized coefficients, a matrix equation is solved
which has the form
\begin{equation}
\Delta_{M(N)}\cdot\boldsymbol{R}_{M(N)}^{(i)\pm}=\boldsymbol{V}_{M(N)}^{(i)\pm},\label{eq:A04}
\end{equation}
where $\Delta$ is the characteristic matrix of the system of equations
from the boundary conditions at the interfaces, $\boldsymbol{R}^{(i)\pm}$
is the column of the generalized coefficients $R_{M(N)}^{\pm(ij)\pm}$
and $\boldsymbol{V}^{(i)\pm}$ is the free term vector whose terms
are given by the homogeneous part of the Green's tensor. 

We will consider in more detail the case where a 2D material, MoS\textsubscript{2},
is sandwiched between two planar half-spaces with dielectric permittivities
$\varepsilon_{1}$ and $\varepsilon_{2}$. The $z$-direction is perpendicular
to the boundary between the two half-spaces\cite{Chew1995,Novotny2012}.
Using Eq.~\eqref{eq:A01} the Green's tensor has the form \begin{subequations}\label{eq:A05}
\begin{equation}
\mathbf{\mathfrak{G}}^{(11)}(\mathbf{r},\mathbf{s},\omega)=\mathfrak{G}_{h}^{(11)}(\mathbf{r},\mathbf{s},\omega)+\mathfrak{G}_{s}^{(11)}(\mathbf{r},\mathbf{s},\omega),\label{eq:A05a}
\end{equation}
\begin{equation}
\mathbf{\mathfrak{G}}^{(21)}(\mathbf{r},\mathbf{s},\omega)=\mathfrak{G}_{s}^{(21)}(\mathbf{r},\mathbf{s},\omega),\label{eq:A05b}
\end{equation}
\end{subequations}where the first of the two labels in the superscript
$(i1)$ denotes the field point, while the second denotes the source
point. The scattering terms have the following expression\begin{widetext}\begin{subequations}\label{eq:A06}

\begin{equation}
\mathfrak{G}_{s}^{(11)}(\mathbf{r},\mathbf{s},\omega)=\frac{i}{8\pi^{2}}\sum_{K}\int\text{d}^{2}k_{s}\frac{1}{k_{z1}k_{s}^{2}}R_{K}^{+11-}\mathbf{K}(k_{s},k_{z1},\mathbf{r})\otimes\mathbf{K}^{*}(k_{s},-k_{z1},\mathbf{s})\label{eq:A06a}
\end{equation}
\begin{equation}
\mathfrak{G}_{s}^{(21)}(\mathbf{r},\mathbf{s},\omega)=\frac{i}{8\pi^{2}}\sum_{K}\int\text{d}^{2}k_{s}\frac{1}{k_{z1}k_{s}^{2}}R_{K}^{-21-}\mathbf{K}(k_{s},-k_{z2},\mathbf{r})\otimes\mathbf{K}^{*}(k_{s},-k_{z1},\mathbf{s})\label{eq:A06b}
\end{equation}
\end{subequations}where $k_{s}=\sqrt{k_{i}^{2}-k_{zi}^{2}}$ is the
in-plane propagation constant, $k_{zi}$ is the perpendicular propagation
constant in medium $i$, and $k_{i}=\frac{\omega}{c}\sqrt{\varepsilon_{i}}$
is the wavenumber in medium $i$ ($i=1,2$). The above expressions
involve a summation over $\mathbf{K}$ which represents $\mathbf{M}$
and $\mathbf{N}$, the transverse electric (TE) and transverse magnetic
(TM) modes, respectively. 

Imposing the continuity conditions, Eq.~\eqref{eq:A05}, at the boundary
between the two half spaces, $z=0$, we obtain the generalized Fresnel
coefficients, which have the form \cite{Hanson2013,Nikitin2013},\begin{subequations}\label{eq:A07}
\begin{equation}
R_{M}^{11}=\frac{k_{z1}-k_{z2}-2\alpha k_{0}}{k_{z1}+k_{z2}+2\alpha k_{0}},\quad R_{N}^{11}=\frac{k_{2}^{2}k_{z1}-k_{1}^{2}k_{z2}+2\alpha k_{0}k_{z1}k_{z2}}{k_{2}^{2}k_{z1}+k_{1}^{2}k_{z2}+2\alpha k_{0}k_{z1}k_{z2}}\label{eq:A07a}
\end{equation}
\begin{equation}
R_{M}^{21}=\frac{2k_{z1}}{k_{z1}+k_{z2}+2\alpha k_{0}},\quad R_{N}^{21}=\frac{2k_{1}k_{2}k_{z1}}{k_{2}^{2}k_{z1}+k_{1}^{2}k_{z2}+2\alpha k_{0}k_{z1}k_{z2}},\label{eq:A07b}
\end{equation}
\end{subequations}where $\alpha=2\pi\sigma/c$.\end{widetext} 

\bibliographystyle{prsty}

\end{document}